\documentclass[reprint, amsmath, amssymb, aps, pre]{revtex4-2}

\usepackage{graphicx} 
\usepackage{array}
\usepackage{dcolumn} 
\usepackage{bm}      
\usepackage{hyperref} 
\usepackage{xcolor}   
\usepackage{caption}

\makeatletter
\providecommand{\href@noop}[2]{#2}
\makeatother

\hypersetup{
    colorlinks=true,
    linkcolor=blue,
    filecolor=magenta,      
    urlcolor=blue,
    citecolor=blue,
}

\begin{document}

\title{Noise-Induced Transitions and Coherence Resonance in a 5D Conductance-Based Neuronal Model}

\author{Yefan Wu}
\email{wuyefan718@gmail.com}
\email{yewu0336@uni.sydney.edu.au} 
\affiliation{Department of Mathematics and Statistics, The University of Sydney, NSW 2006, Australia}

\begin{abstract}
Intrinsic channel noise is an important source of variability in neuronal dynamics, but its state-dependent and boundary-constrained nature can be difficult to represent numerically. Here, we investigate the effects of bounded multiplicative noise applied to the slow M-current gating variable in a five-dimensional conductance-based model of a CA1 pyramidal neuron. We employ a full-truncation semi-implicit Euler scheme to control boundary violations and perform Monte Carlo parameter sweeps, time-step checks, burst-detection sensitivity analyses, and conductance perturbations. In subthreshold regimes, noise induces firing and produces an intermediate-noise minimum in the coefficient of variation, consistent with coherence resonance. In the deep subthreshold regime, the burst rate exhibits moderate Arrhenius-like scaling over a restricted low-noise interval, supporting an interpretation in terms of noise-activated escape. Near the numerically identified onset of sustained oscillations, temporal coherence is particularly sensitive to noise intensity. In the suprathreshold regime, strong multiplicative noise accelerates firing through premature stochastic exits from the hyperpolarized recovery branch, although this behavior does not follow classical Kramers scaling. Control simulations using clipped additive Gaussian noise produce qualitatively different dynamics, indicating that the state dependence and boundary behavior of the diffusion term materially affect the observed transitions. The main trends remain robust under the tested variations in numerical resolution, burst-detection threshold, slow timescale, and M-current conductance. These results characterize model-specific effects of boundary-constrained channel noise and motivate future comparisons with discrete-state ion-channel models.
\end{abstract}

\maketitle

\section{Introduction}
\label{sec:intro}

Stochastic fluctuations are ubiquitous in complex dynamical systems, traditionally viewed merely as a source of disorder and entropy. However, nonlinear systems possess the notable ability to harness noise constructively, manifesting in counter-intuitive phenomena such as stochastic resonance (SR) and coherence resonance (CR) \cite{pikovsky1997coherence, gammaitoni1998stochastic, lindner2004effects}. Understanding how noise interacts with nonlinear manifolds and their basins of attraction to orchestrate dynamical transitions is thus important across nonlinear dynamics, statistical mechanics, and computational neuroscience \cite{yoon2024tracking, freidlin2012random}.

Despite extensive research on noise-induced transitions, the majority of theoretical paradigms rely on highly simplified, low-dimensional approximations subjected to unbounded, additive Gaussian noise \cite{yamakou2020optimal}. Furthermore, many excitable systems inherently possess a fast-slow functional structure \cite{wechselberger2020geometric, berglund2006noise}, where noise uniquely interacts with the attracting branches of critical manifolds. While recent advances have begun to explore the effects of multiplicative noise in higher-dimensional biophysical models, such as cardiac action potentials \cite{bashkirtseva2025analysis}, these studies often overlook the mathematical complexities of bounded domains. In biological neurons, intrinsic noise primarily arises from the stochastic gating of finite ion channels \cite{faisal2008noise}. This channel noise is inherently multiplicative and state-dependent, vanishing at the boundaries of the probability domain $[0,1]$ to satisfy Feller boundary conditions \cite{goldwyn2011what, yu2022effects}. The numerical integration of such bounded stochastic differential equations (SDEs) requires specialized mathematical treatments to preserve the probability domain. To accurately capture the behavior of the 5D manifold under extreme multiplicative noise, advanced implicit numerical frameworks are essential \cite{higham2005convergence, lord2010comparison, hutzenthaler2020perturbation}.

To bridge this gap, we investigate a realistic 5D Hodgkin-Huxley-type conductance model of a CA1 cortical pacemaker \cite{golomb2006contribution}. We specifically introduce a bounded, state-dependent multiplicative noise to the slow M-type potassium current ($z$ gating variable). While macroscopic models typically assume weak fluctuations, the effective noise intensity scales inversely with the square root of the local ion channel population ($\sigma_z \propto 1/\sqrt{N}$). Therefore, extreme multiplicative noise is not a mathematical artifact, but a direct macroscopic proxy for channelopathies such as benign familial neonatal convulsions (BFNC), where pathogenic excitability emerges in dendritic microdomains with severely restricted channel numbers ($N < 100$) \cite{mittal2022heterogeneous}. Although modeling such extreme low-$N$ limits ultimately points toward exact discrete Markov jump processes \cite{fox1994emergent, gillespie1977exact}, the bounded Feller diffusion provides a mathematically consistent and computationally tractable framework to capture the macroscopic active boundary accumulations. While discrete simulations offer exactitude at the single-channel level, the continuous Feller SDE framework enables the analytical deployment of Fokker-Planck reductions and the mapping of smooth phase-space manifolds, providing a bridge between microscopic stochasticity and macroscopic dynamical systems theory. As a slow inhibitory conductance, the M-current dictates the timescale of bursting and recovery \cite{gutkin1998dynamics}. Because random fluctuations in slow variables can modulate the excitability of fast variables, this setup provides an ideal testbed to explore how physically constrained multiplicative noise manipulates the neural state space near distinct bifurcation boundaries.

In this paper, we employ a full-truncation semi-implicit Euler scheme to characterize the dynamical landscape of the 5D manifold. We identify a triphasic stochastic response dictated by the underlying bifurcation structure \cite{ermentrout2010mathematical}. First, in the deep sub-threshold regime, multiplicative noise acts constructively, triggering stochastic awakening from deterministic quiescence via Kramers escape \cite{hanggi1990reaction}. Second, at the subcritical Hopf boundary, the system exhibits enhanced sensitivity where intermediate noise leverages the proximity to the limit cycle, inducing robust coherence resonance \cite{pikovsky1997coherence}. In addition, in the supra-threshold rhythmic regime, we observe a dynamical shift toward noise-accelerated Kramers escape. Under multiplicative noise, the Feller boundary shapes the probability landscape near the biophysical limits, facilitating premature stochastic exits from the hyperpolarized slow manifold and a noise-accelerated escape that differs from classical additive theories. To test whether the Feller boundary geometry is structurally important for this mechanism, we perform controlled knockout experiments replacing the Feller-type noise with unbounded Gaussian noise. This intervention abolishes bursting in both the quiescent and oscillatory regimes, indicating that state-dependent boundary preservation, rather than merely the presence of noise fluctuations, is required for the observed noise-accelerated bursting dynamics. Finally, conductance perturbation experiments confirm the biological robustness of this noise-induced dynamical transition, delineating the limits of temporal coherence in high-dimensional excitable media.

\section{Model and Methods}
\label{sec:methods}

\subsection{Deterministic 5D Cortical Pacemaker Model}
To provide a biophysically grounded exploration of the high-dimensional state space, we adopt a single-compartment conductance-based model of a CA1 pyramidal neuron, as developed by Golomb \textit{et al.} \cite{golomb2006contribution}. The temporal evolution of the membrane potential $V$ (in mV) is governed by the current balance equation:
\begin{equation}
C_m \frac{dV}{dt} = - (I_{\mathrm{Na}} + I_{\mathrm{NaP}} + I_{\mathrm{Kdr}} + I_{\mathrm{A}} + I_{\mathrm{M}} + I_{\mathrm{leak}}) + I_{\mathrm{app}},
\label{eq:voltage_balance}
\end{equation}
where $C_m = 1.0 \, \mu\text{F/cm}^2$ is the membrane capacitance, and $I_{\mathrm{app}}$ serves as the primary bifurcation parameter. 

Each ionic current in the model corresponds to a specific mechanism observed in CA1 neurons. Specifically, the fast sodium current $I_{\mathrm{Na}}$ initiates the action-potential upstroke. The persistent sodium current $I_{\mathrm{NaP}}$ provides a sustained depolarizing drive and acts as the burst sustainer. The delayed-rectifier potassium current $I_{\mathrm{Kdr}}$ enables rapid repolarization, while the A-type transient potassium current $I_{\mathrm{A}}$ regulates the time of spiking. 

Importantly, the M-type potassium current $I_{\mathrm{M}}$ represents slow adaptation and functions as the burst terminator. Its activation is controlled by the slow gating variable $z$, which forms the focal point of our stochastic analysis. The leak current $I_{\mathrm{L}}$ maintains the resting potential. The mathematical formulations and maximal conductances $g_x$ are summarized in Table~\ref{tab:currents}.

\begin{table}[htpb]
\caption{Formulations and parameters for the 5D CA1 pacemaker model. Maximal conductances ($g_x$) are given in $\text{mS/cm}^2$ and reversal potentials ($V_x$) are in $\text{mV}$.}
\label{tab:currents}
\begin{ruledtabular} 
\begin{tabular}{l l}  
Current & Formulation and Maximal Conductance \\
\hline 
$I_{\mathrm{Na}}$  & $g_{\mathrm{Na}}m_\infty^3(V)h(V - V_{\mathrm{Na}})$; $g_{\mathrm{Na}} = 35$, $V_{\mathrm{Na}} = 55$ \\
$I_{\mathrm{NaP}}$ & $g_{\mathrm{NaP}}p_\infty(V)(V - V_{\mathrm{Na}})$; $g_{\mathrm{NaP}} = 0.25$ \\
$I_{\mathrm{Kdr}}$ & $g_{\mathrm{Kdr}}n^4(V - V_{\mathrm{K}})$; $g_{\mathrm{Kdr}} = 6$, $V_{\mathrm{K}} = -90$ \\
$I_{\mathrm{A}}$   & $g_{\mathrm{A}}a_\infty^3(V)b(V - V_{\mathrm{K}})$; $g_{\mathrm{A}} = 1.4$ \\
$I_{\mathrm{M}}$   & $g_{\mathrm{M}}z(V - V_{\mathrm{K}})$; $g_{\mathrm{M}} = 1.0$ \\
$I_{\mathrm{leak}}$& $g_{\mathrm{leak}}(V - V_{\mathrm{leak}})$; $g_{\mathrm{leak}} = 0.05$, $V_{\mathrm{leak}} = -70$ \\
\end{tabular}
\end{ruledtabular}
\end{table}

The activation variables ($m, p, a$) are treated as instantaneous, taking their steady-state values $x_\infty(V)$. The dynamic gating variables $y \in \{h, n, b\}$ evolve according to $dy/dt = (y_\infty(V) - y)/\tau_y(V)$. The time constants $\tau_h(V)$, $\tau_n(V)$, and the static value $\tau_b = 15.0$ ms follow the exact parametrization in \cite{golomb2006contribution}. Because the macroscopic relaxation of $z$ (introduced below, with $\tau_z = 75.0$ ms) is significantly slower than the kinetics of these fast variables, the 5D model inherently possesses a well-defined fast-slow timescale separation \cite{kuehn2015multiple, wechselberger2020geometric}. This architectural feature defines a critical manifold where the slow variable $z$ governs the system's traversal during inter-burst recovery.

\subsection{State-Dependent Multiplicative Channel Noise}
To investigate the impact of biophysically constrained noise, we introduce a state-dependent multiplicative noise to the slow M-type potassium gating variable $z$. Adopting the Itô interpretation, the stochastic differential equation (SDE) for $z$ is formulated as:
\begin{equation}
dz = \frac{z_\infty(V) - z}{\tau_z} dt + \sigma_z \sqrt{z(1-z)} dW_t,
\label{eq:sde_z}
\end{equation}
where $\tau_z = 75.0$ ms and $dW_t$ represents the standard Wiener process increment. The multiplicative term $\sqrt{z(1-z)}$ ensures that the noise satisfies Feller boundary conditions, dynamically scaling the stochastic fluctuations to zero as $z \to 0$ or $z \to 1$. While this guarantees that the continuous-time trajectory remains within the biophysical probability domain $[0, 1]$ \cite{goldwyn2011what, yu2022effects}, preserving this boundary during discrete numerical integration requires specialized schemes, as detailed in the following subsection.

\subsection{Numerical Integration Scheme and Statistical Measures}
\label{sec:numerical}

The numerical integration of SDEs with Feller boundary conditions presents mathematical challenges. Standard explicit integration schemes (e.g., Euler-Maruyama) cannot reliably guarantee domain positivity in the presence of strong multiplicative noise, which typically leads to domain violations (i.e., $z < 0$ or $z > 1$) and complex-valued diffusion terms. To preserve the biophysical probability domain $z \in [0,1]$ and maintain numerical stability for the highly nonlinear gating kinetics, we employed a full-truncation semi-implicit Euler scheme \cite{higham2005convergence, lord2010comparison}.

In this framework, the membrane potential $V$ is integrated using the explicit forward Euler method, while all gating variables $y \in \{h, n, b\}$ are updated implicitly to stabilize the rapid transient kinetics \cite{rush1978practical, ermentrout2010mathematical}:
\begin{equation}
y_{t+\Delta t} = \frac{y_t + y_\infty(V_t) \frac{\Delta t}{\tau_y(V_t)}}{1 + \frac{\Delta t}{\tau_y(V_t)}}.
\end{equation}

For the stochastic M-type gating variable $z$, we adopt the full-truncation approach \cite{lord2010comparison, hutzenthaler2020perturbation} to evaluate the Feller diffusion term. Let $\tilde{z} = \max(\min(z_t, 1), 0)$ denote the domain-restricted effective state \cite{lord2010comparison}. This explicit truncation is applied exclusively to the diffusion term to prevent undefined mathematical operations, while the drift term evolves the true continuous state $z_t$ within the implicit framework. The semi-implicit update rule is formulated as:
\begin{equation}
z_{t+\Delta t} = \frac{z_t + z_\infty(V_t) \frac{\Delta t}{\tau_z} + \sigma_z \sqrt{\tilde{z}(1 - \tilde{z})} \Delta W_t}{1 + \frac{\Delta t}{\tau_z}},
\end{equation}
where $\Delta W_t \sim N(0, \Delta t)$ is the standard Wiener increment. Importantly, we do not impose any secondary, artificial reflecting or capping boundaries on the updated state $z_{t+\Delta t}$. Because the forward Wiener increment is an unbounded Gaussian, infinitesimally small out-of-bounds excursions can theoretically occur. However, under the full-truncation scheme, any such excursion immediately zeroes out the diffusion term in the subsequent step ($\tilde{z}=0$ or $\tilde{z}=1$), allowing the implicitly integrated, strongly attracting drift to naturally and smoothly pull the trajectory back into the valid biophysical domain $[0,1]$ \cite{lord2010comparison}. This semi-implicit treatment conserves the probability flow at the physical boundaries without introducing ad-hoc absorption traps. To capture the bifurcation boundary and ensure convergence, the system was simulated with a fine temporal resolution of $\Delta t = 0.01$ ms.

Furthermore, to validate numerical convergence under extreme multiplicative shocks ($\sigma_z > 10^{-1}$), we empirically verified our scheme by halving the time step ($\Delta t = 0.005$ ms). At a representative test point ($I_{\mathrm{app}} = 0.45$, $\sigma_z = 0.1$, 50 trials), the mean burst rate and CV each changed by less than 2\% and 5\% respectively, well within acceptable tolerance. This refined resolution yielded quantitatively consistent macroscopic phase diagrams, confirming the convergence of our semi-implicit approach.

While more advanced boundary-preserving schemes, such as those employing projection or reflection methods, could potentially be developed, our extensive timestep refinement tests and the good agreement with the analytical Fokker-Planck solution in the valid regime (Fig.~\ref{fig:fpe_verification}) confirm that our full-truncation semi-implicit scheme is robust and sufficient to capture the qualitative and quantitative dynamics reported herein.

For statistical analysis, independent trials driven by uncorrelated Wiener processes were simulated. To eliminate transient effects and ensure statistical stationarity, all simulations were initialized from their respective deterministic steady-states or limit cycles, and an initial burn-in period of 500 ms was discarded before any temporal metrics or probability densities were computed. To minimize spurious correlations across noise intensities, each trial was seeded with an independent random state, ensuring statistical independence within and across simulation conditions. All reported statistics are computed over 50 independent realizations per parameter set, with standard errors of the mean (SEM) as error bars.

\begin{figure*}[t]
    \centering
    \includegraphics[width=0.9\textwidth, trim=0 0 0 0.8cm, clip]{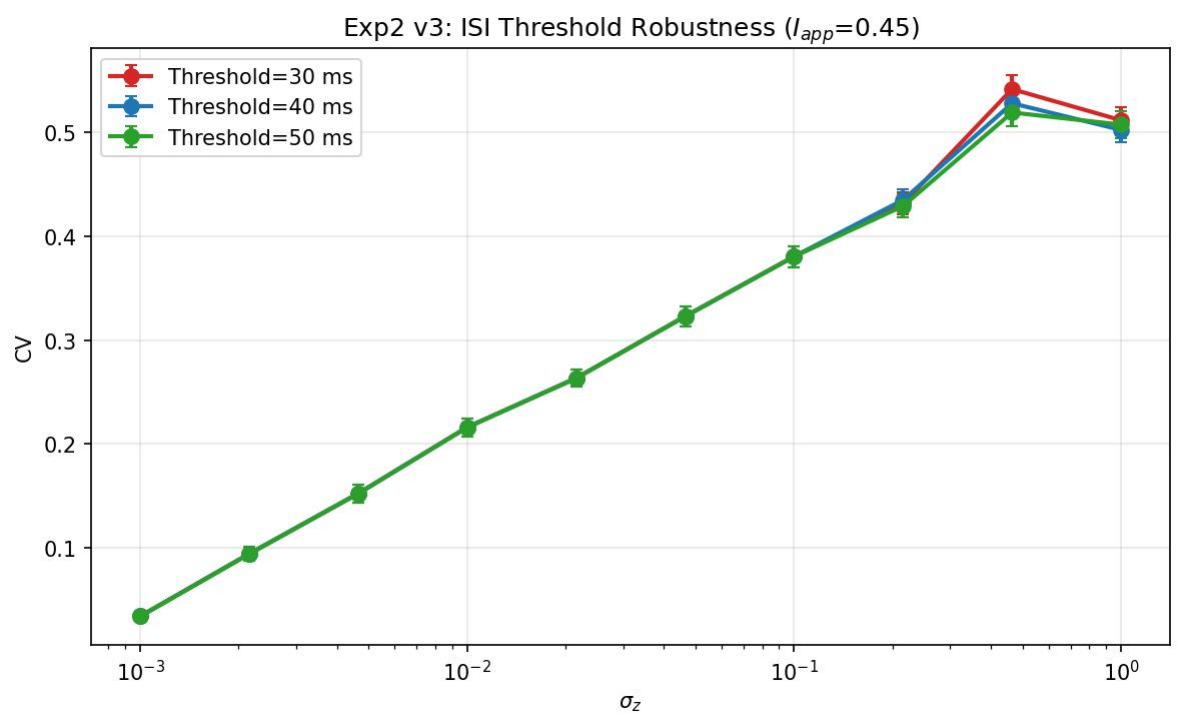}
    \caption{Robustness of the CV curves to variations in the burst detection threshold. CV as a function of noise intensity $\sigma_z$ evaluated at three different inter-spike interval thresholds ($T_{\mathrm{threshold}} = 30, 40, 50$ ms) across the four dynamical regimes. Computed from 50 independent trials per point with per-trial seeding. The qualitative shape of the CV valleys and the triphasic transition structure remain invariant across all three threshold values, confirming that the observed coherence properties are intrinsic features of the system rather than artifacts of the burst detection algorithm.}
    \label{fig:threshold_robustness}
\end{figure*}

To parse the time series, a new burst was computationally identified when the inter-spike interval (ISI) exceeded a threshold of 40 ms, effectively distinguishing macroscopic inter-burst intervals from deterministic high-frequency intra-burst spikes. This threshold was chosen to cleanly decouple the fast intra-burst spiking timescale (driven by rapid $\text{Na}^+$ and $\text{K}^+$ kinetics) from the slow inter-burst recovery timescale (governed by the macroscopic relaxation of the M-current), consistent with standard burst detection criteria in CA1 models \cite{golomb2006contribution}. Furthermore, we systematically verified that the qualitative shape of the CV curves and the observed triphasic transitions remain robust to variations in this burst detection threshold (testing $T_{\mathrm{threshold}} \in[30, 50]$ ms; see Fig.~\ref{fig:threshold_robustness}). Defining $T_i = t_{i+1} - t_i$ as the time interval between the first spikes of two consecutive bursts, the CV is given by:
\begin{equation}
CV = \frac{\sqrt{\vphantom{\langle T \rangle} \langle T^2 \rangle - \langle T \rangle^2}}{\langle T \rangle},
\end{equation}
where $\langle \cdot \rangle$ denotes the temporal average over a sufficiently long simulation epoch. A minimum threshold of $N_{\mathrm{bursts}} \ge 3$ was required to calculate the CV, effectively filtering out undefined statistical variances in the deep sub-threshold regime where the system remains in deterministic quiescence.

\section{Results}
\label{sec:results}

\begin{figure*}[t]
    \centering
    \begin{minipage}{0.48\textwidth}
        \centering
        \includegraphics[width=\linewidth]{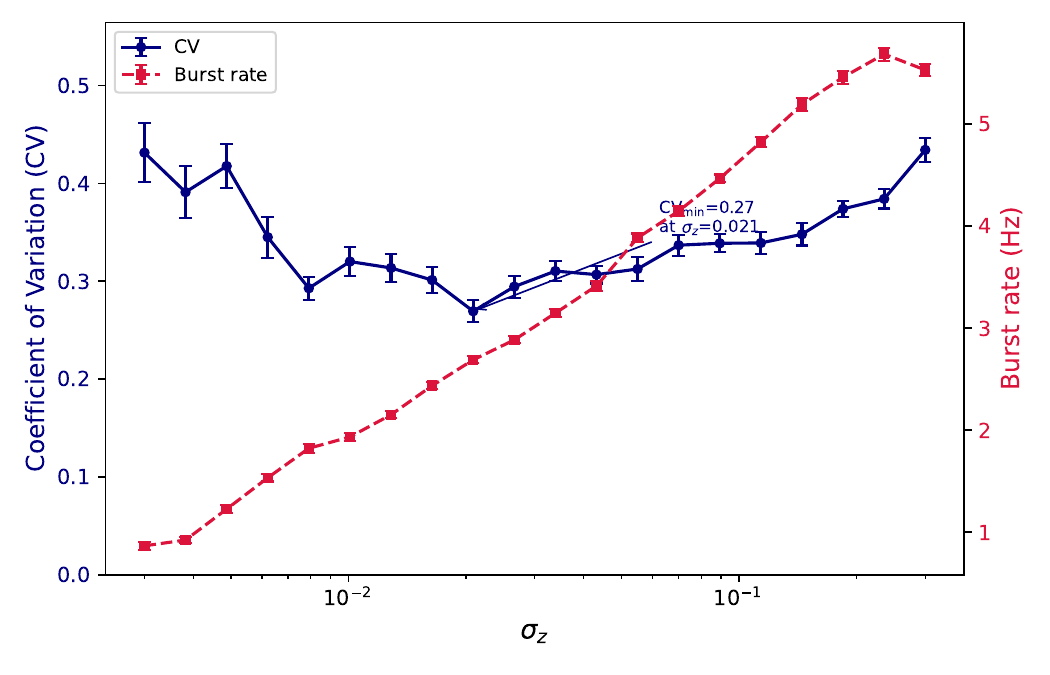}
        \caption*{(a) Deep sub-threshold ($I_{app} = 0.35$) }
    \end{minipage}\hfill
    \begin{minipage}{0.48\textwidth}
        \centering
        \includegraphics[width=\linewidth]{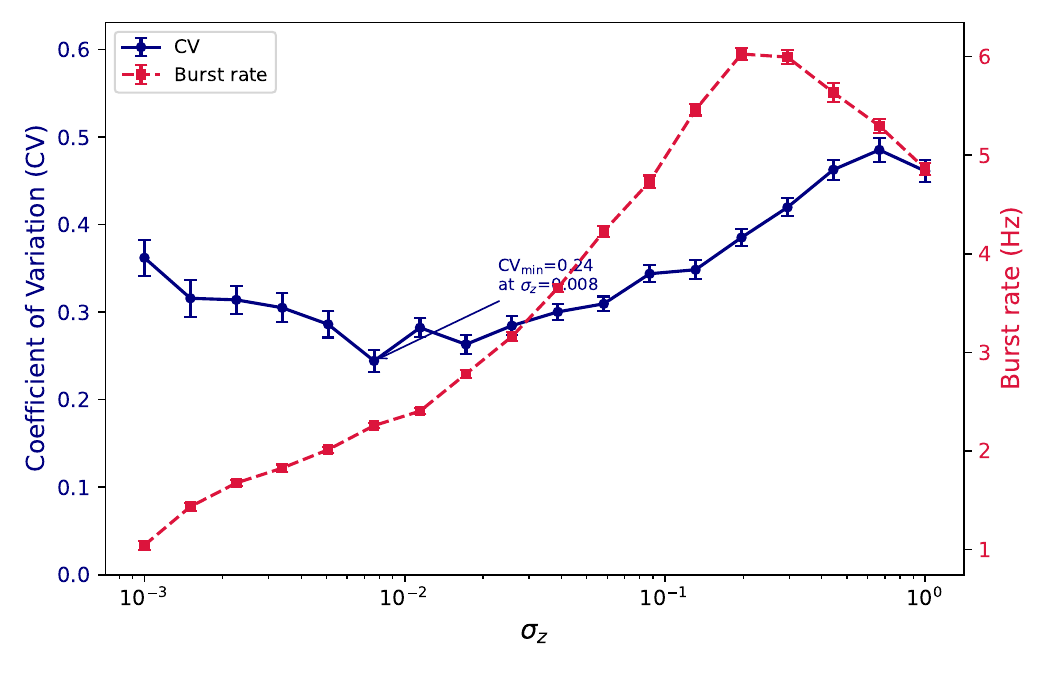}
        \caption*{(b) Subcritical edge ($I_{app} = 0.39$) }
    \end{minipage}
    
    \vspace{0.5cm} 
    
    \begin{minipage}{0.48\textwidth}
        \centering
        \includegraphics[width=\linewidth]{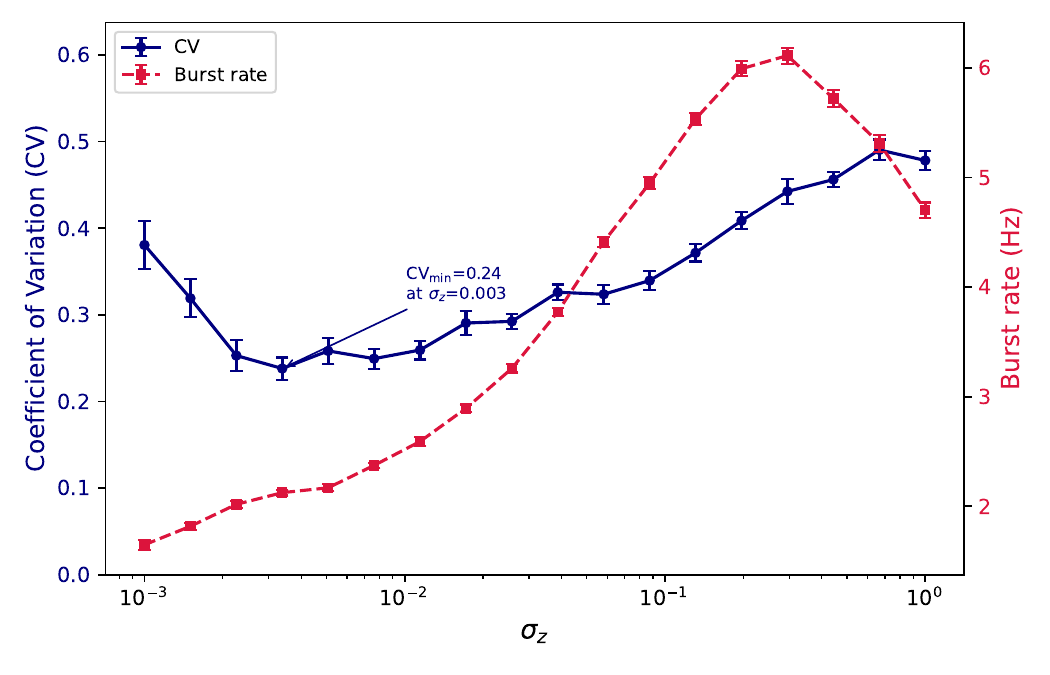}
        \caption*{(c) Critical boundary ($I_{app} = 0.3955$) }
    \end{minipage}\hfill
    \begin{minipage}{0.48\textwidth}
        \centering
        \includegraphics[width=\linewidth]{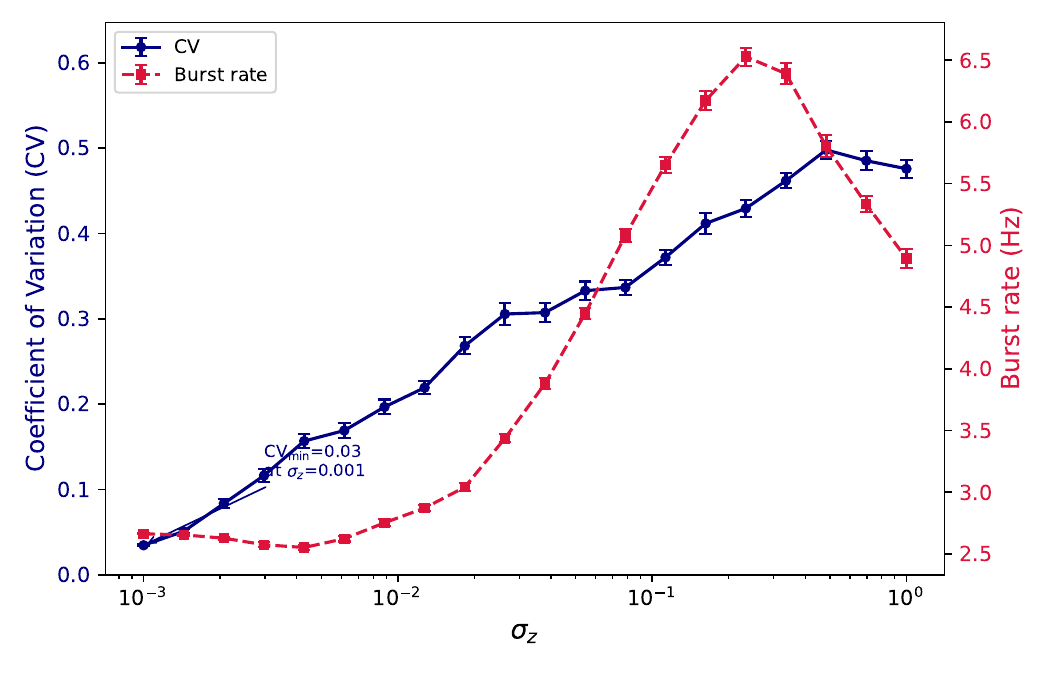}
        \caption*{(d) Supra-threshold pacemaker ($I_{app} = 0.45$) }
    \end{minipage}
    
    \caption{The Triphasic Landscape of Noise-Induced Transitions in the 5D Cortical Manifold. (a) Deep sub-threshold regime ($I_{app} = 0.35$): Multiplicative noise triggers stochastic awakening from deterministic quiescence. (b) Sub-threshold resonance ($I_{app} = 0.39$): The system exhibits a broad coherence valley, maintaining stable noise-driven pacing. (c) Critical boundary ($I_{app} = 0.3955$): Proximity to the subcritical Hopf bifurcation induces a sensitive, sharply tuned coherence resonance (CR) conducive to rhythmic pacing. (d) Supra-threshold rhythmic regime ($I_{app} = 0.45$): Multiplicative noise monotonically degrades deterministic coherence. Under the proposed semi-implicit framework, large noise intensities ($\sigma_z > 10^{-1}$) drive noise-accelerated Kramers escape across all regimes, pushing the system into high-frequency, jittered stochastic bursting.}
    \label{fig:grand_landscape}
\end{figure*}

\begin{figure*}[t]
    \centering
    \begin{minipage}{0.48\textwidth}
        \centering
        \includegraphics[width=\linewidth]{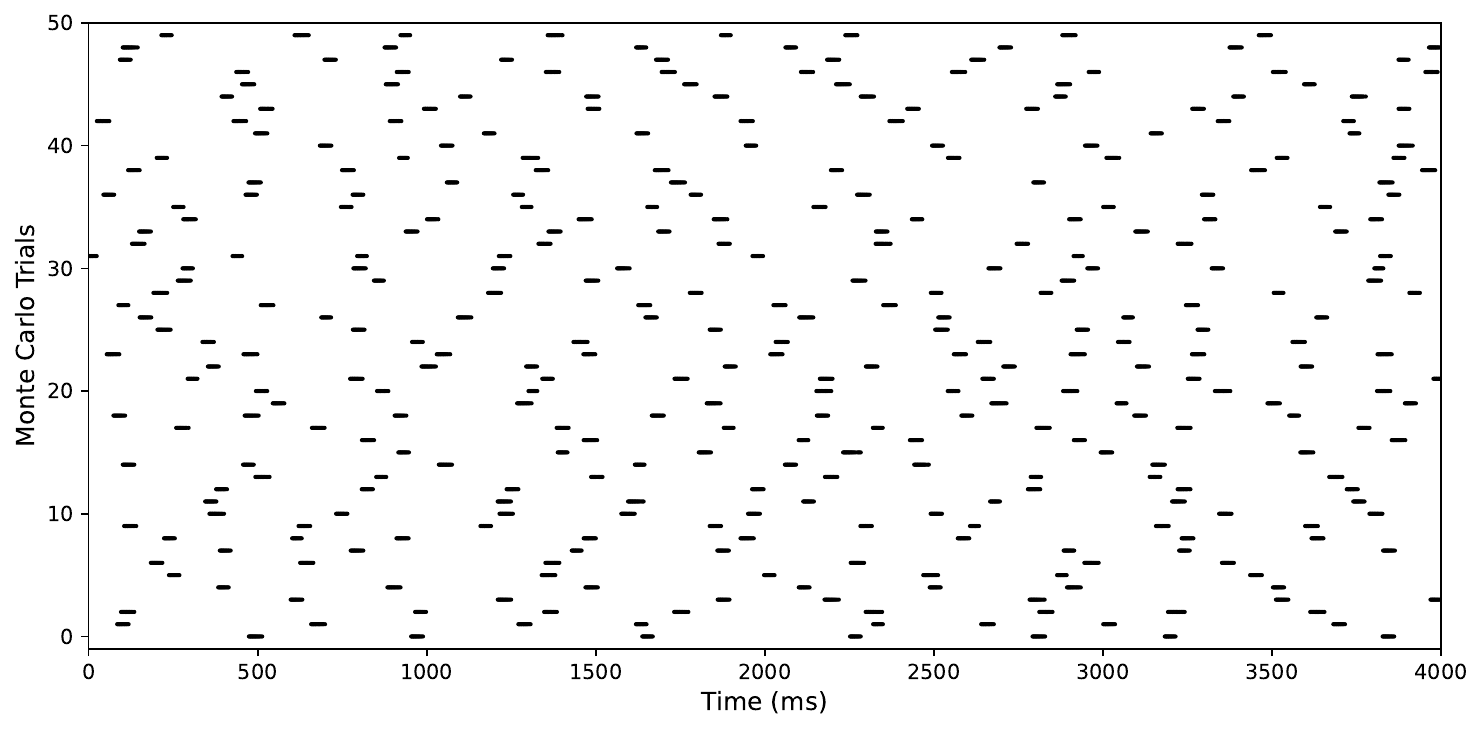}
        \caption*{(a) Awakening ($0.35$)}
    \end{minipage}\hfill 
    \begin{minipage}{0.48\textwidth}
        \centering
        \includegraphics[width=\linewidth]{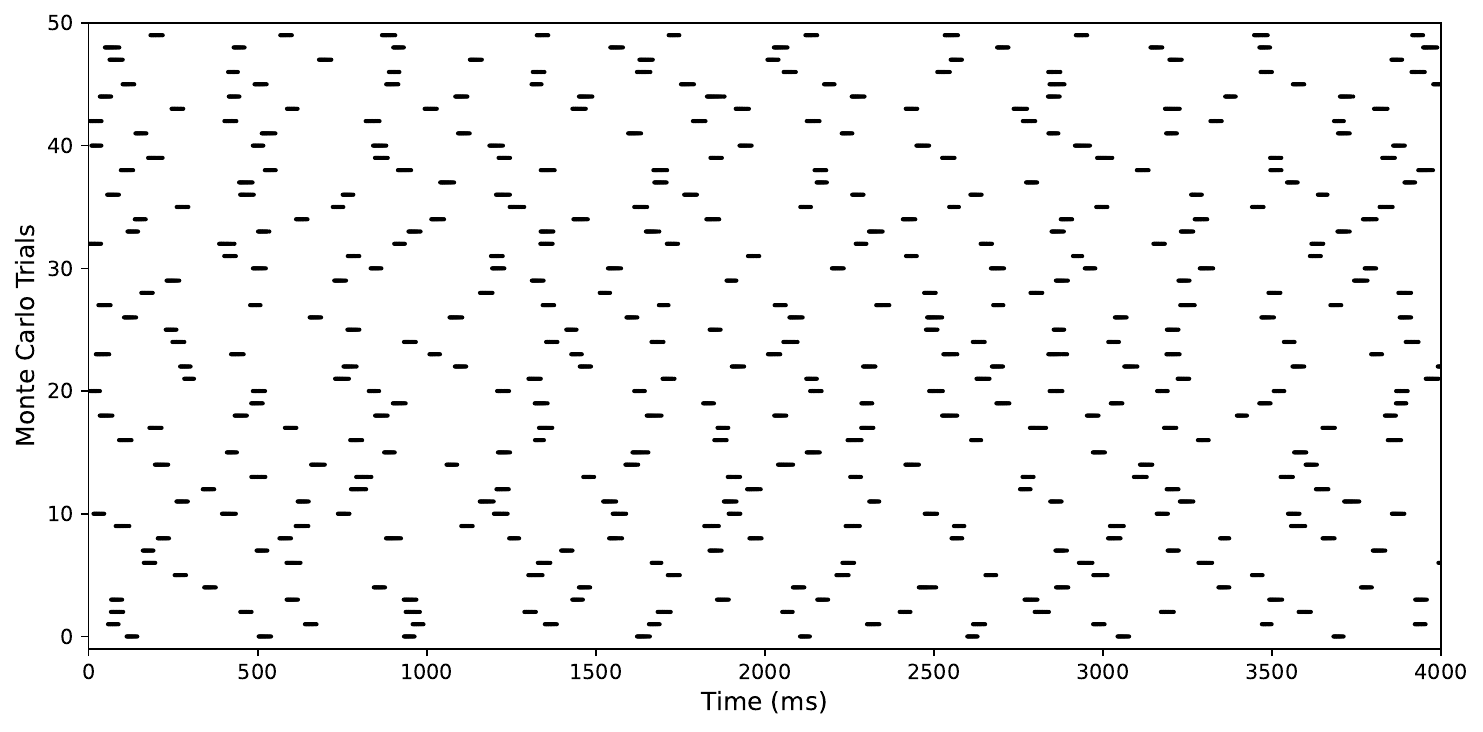}
        \caption*{(b) Coherence ($0.39$)}
    \end{minipage}
    
    \vspace{0.5cm}
    
    \begin{minipage}{0.48\textwidth}
        \centering
        \includegraphics[width=\linewidth]{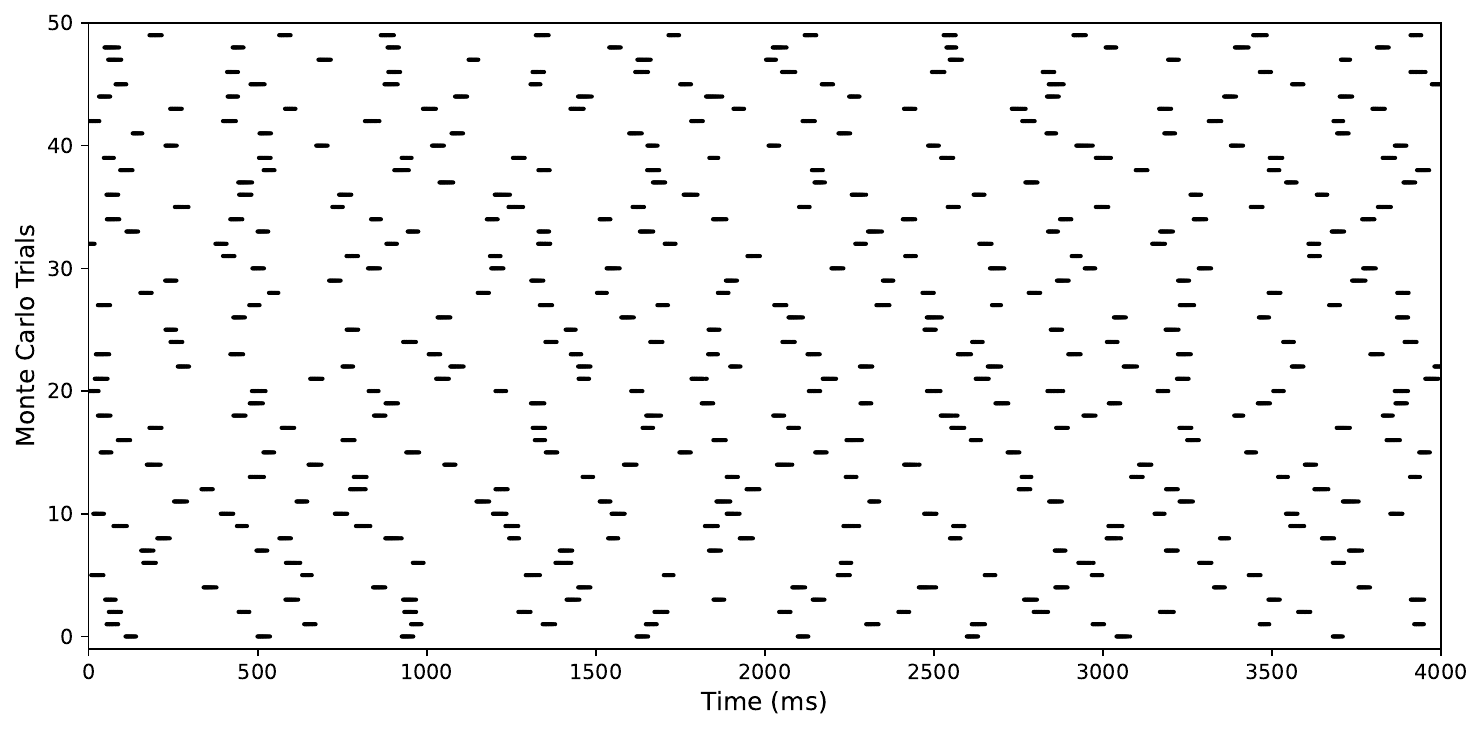}
        \caption*{(c) Sensitivity ($0.3955$)}
    \end{minipage}\hfill
    \begin{minipage}{0.48\textwidth}
        \centering
        \includegraphics[width=\linewidth]{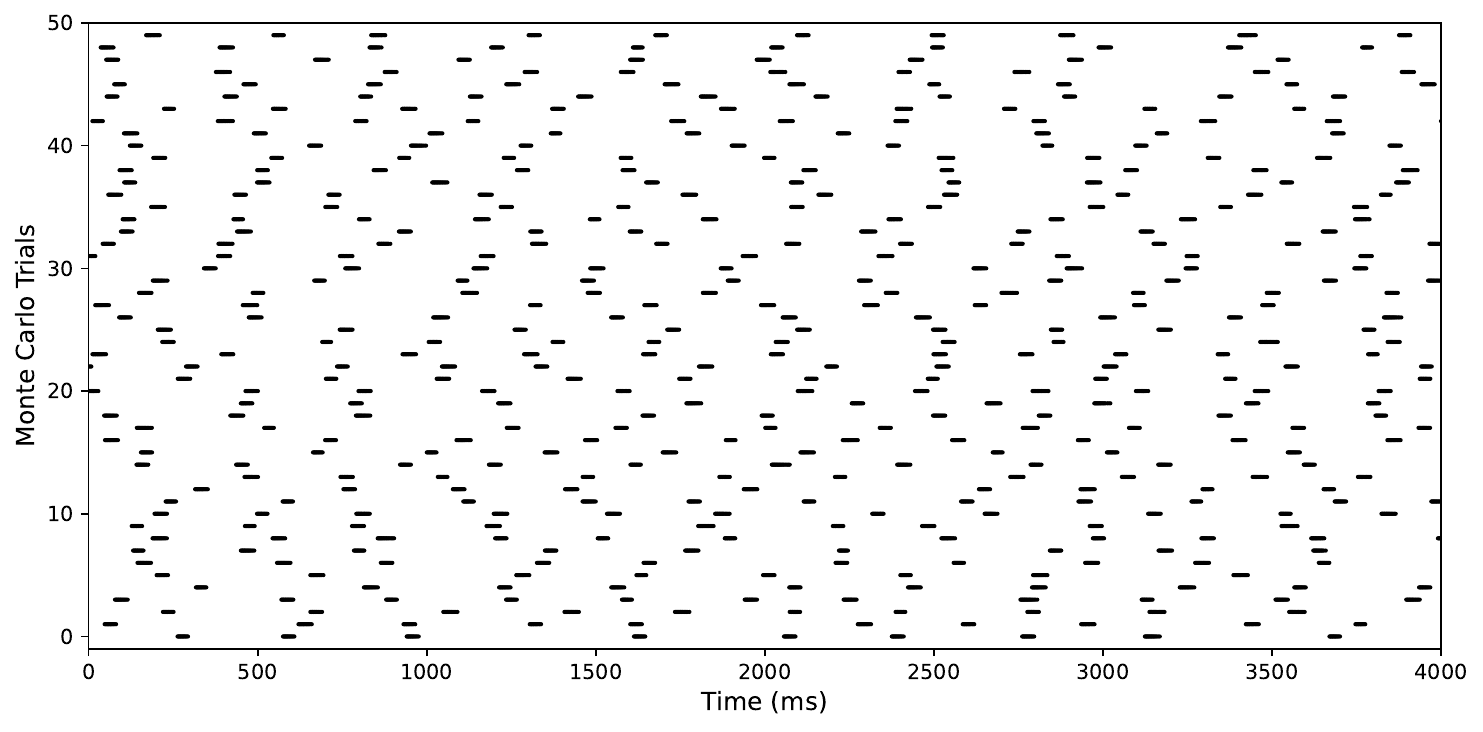}
        \caption*{(d) Stress ($0.45$)}
    \end{minipage}
    
    \caption{Spiking Raster Plots across Four Dynamical Regimes at an Intermediate Noise Intensity ($\sigma_z = 0.01$). Each panel displays 50 independent Monte Carlo trials over a 4000 ms window. (a) Sparse, noise-driven bursts in the deep sub-threshold regime where the stochastic energy only occasionally overcomes the saddle-node barrier. (b) Coherence resonance just below the Hopf bifurcation: while inter-trial phase slipping creates diagonal drift, each individual row exhibits regular horizontal spike intervals. (c) Secondary stochastic phase-locking at the critical boundary: the noise stabilizes the nascent limit cycle, balancing phase jittering with resonant re-entrainment. (d) Jittered spiking patterns in the supra-threshold healthy pacemaker; as confirmed later, this represents a noise-driven stochastic limit cycle rather than deterministic chaos, indicating that multiplicative noise disrupts intrinsic rhythmicity.}
    \label{fig:rasters}
\end{figure*}

To systematically investigate the dynamical influence of multiplicative channel noise, we must first delineate the underlying deterministic bifurcation structure of the 5D CA1 pacemaker model. We conducted a high-resolution parameter sweep of the applied current ($I_{\mathrm{app}}$) in the absence of noise ($\sigma_z = 0$), identifying the exact critical boundary where the system transitions from a quiescent stable fixed point to a sustained limit-cycle bursting state.

Through micro-sweeps with a precision of $\Delta I_{\mathrm{app}} = 10^{-4} \, \mu\mathrm{A/cm}^2$, we located the deterministic oscillation threshold at approximately $I_{\mathrm{app}} \approx 0.395$. Based on this deterministic bifurcation structure, we selected four dynamical coordinates for our stochastic analysis:

\begin{itemize}
    \item $I_{\mathrm{app}} = 0.35$ (Sub-threshold): An inhibited state with a deep potential well, used to observe noise-induced awakening via barrier escape.
    \item $I_{\mathrm{app}} = 0.39$ (Subcritical Edge): Located marginally below the bifurcation boundary. The system has high excitability but ultimately decays to rest, providing conditions for coherence resonance.
    \item $I_{\mathrm{app}} = 0.3955$ (Near Criticality): Positioned immediately across the bifurcation line. A nascent, fragile limit cycle is born, allowing us to probe boundary-specific phase-space sensitivities and multi-phase transitions.
    \item $I_{\mathrm{app}} = 0.45$ (Supra-threshold): A self-sustained periodic pacemaker regime, serving as a control for the effects of multiplicative noise.
\end{itemize}

\begin{figure*}[htbp]
    \centering
    \includegraphics[width=0.95\textwidth, trim=0 0 0 2.62cm, clip]{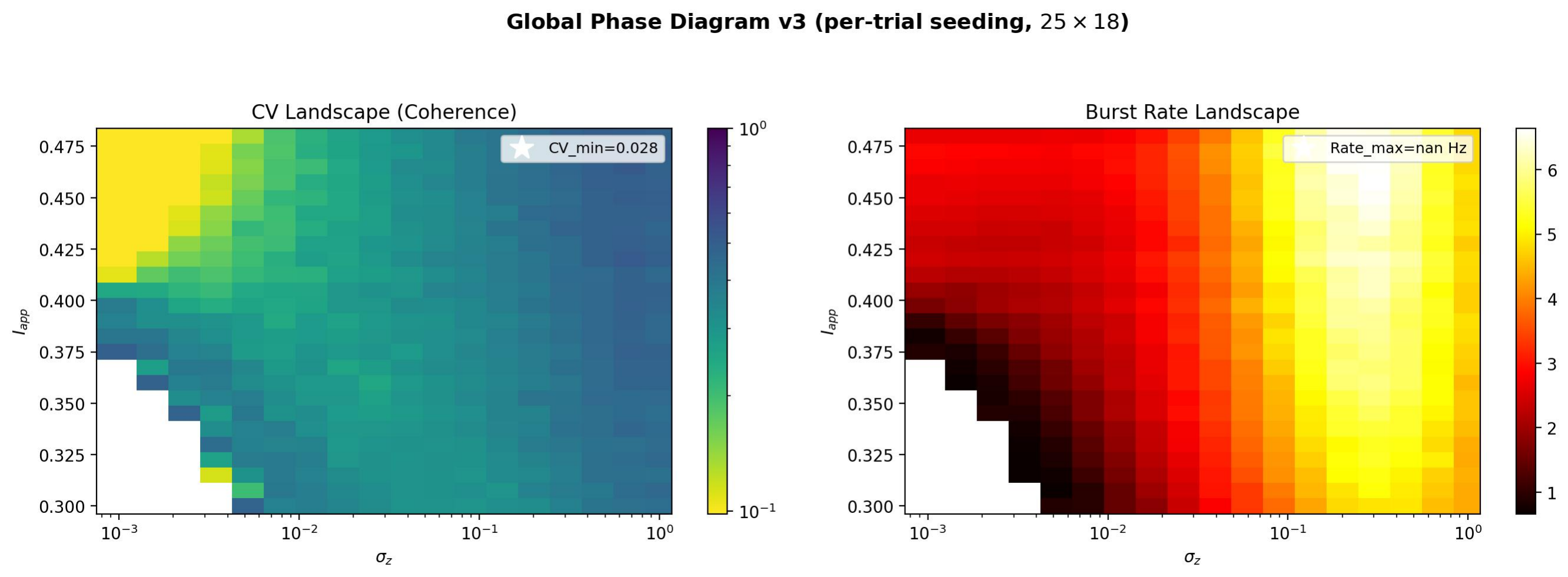} 
    \caption{Global phase diagram of the 5D cortical manifold under Feller-type multiplicative noise. (Left) The Coefficient of Variation (CV) landscape reveals a distinct, leftward-tilting ``valley of coherence'' (dark blue region) near the subcritical Hopf bifurcation boundary. This geometric tilt indicates that deeper sub-threshold states require exponentially larger Feller diffusion to achieve optimal rhythmic pacing, consistent with the Arrhenius-type scaling of barrier crossing in high-dimensional potentials. (Right) The macroscopic burst rate landscape demonstrates the consistent onset of noise-accelerated Kramers escape. At large noise intensities ($\sigma_z > 10^{-1}$), the system transitions into a high-frequency bursting regime (bright wall), overriding the deterministic baseline dynamics and perturbing the basin of attraction of the hyperpolarized slow manifold. This high-resolution phase diagram was computed from 25 $\times$ 18 grid points with 50 independent trials per point and per-trial independent seeding to eliminate spurious correlations across noise intensities.}
    \label{fig:heatmap}
\end{figure*}

With this deterministic skeleton established, we now introduce the bounded multiplicative noise ($\sigma_z > 0$) to trace the stochastic evolution of the 5D manifold. In the following subsection, we analyze the micro-dynamics of the system along these four $I_{\mathrm{app}}$ coordinates, revealing a triphasic transition driven by noise-accelerated Kramers escape across the fast-slow manifold. This triphasic evolution is not a local artifact but a global property, as evidenced by the high-frequency ``wall'' in the macroscopic landscape (Fig.~\ref{fig:heatmap}), where large fluctuations consistently override the deterministic boundaries.

\subsection{Regime I: Stochastic Awakening via Kramers Escape ($I_{\mathrm{app}} = 0.35$)}

In the deep sub-threshold regime ($I_{\mathrm{app}} = 0.35$), the deterministic system is deeply trapped within a stable quiescent fixed point, failing to elicit any action potentials. However, the introduction of state-dependent multiplicative noise to the slow M-current variable $z$ significantly reshapes this dormant energy landscape. 

As illustrated in Fig.~\ref{fig:grand_landscape}(a), the system undergoes a well-defined process of stochastic awakening. At minimal noise intensities ($\sigma_z < 10^{-3}$), the bounded Feller fluctuations are insufficient to overcome the deterministic activation barrier, and the neuron remains silent (burst rate effectively 0 Hz). As $\sigma_z$ increases, noise-driven excursions across the saddle-node separatrix become increasingly frequent, effectively kicking the system out of the quiescent well. This dynamical transition is characterized by a marked increase in the burst rate, climbing from near-zero to a peak of $\sim 5.7$ Hz at $\sigma_z \approx 0.235$ before declining under extreme noise, consistent with classical Kramers escape rate theory for noise-activated barrier crossing in the low-to-moderate noise regime \cite{hanggi1990reaction}. 

Notably, the temporal regularity of these noise-induced bursts is not purely random; it exhibits a broad coherence valley. As noise facilitates more consistent barrier crossings, the Coefficient of Variation (CV) steeply declines, reaching a broad local minimum ($CV \approx 0.27$ at $\sigma_z \approx 2 \times 10^{-2}$) before rising again when fluctuations begin to jitter the inter-burst intervals. This wideband resonance is characteristic of the saddle-node bifurcation's slow manifold, indicating that bounded multiplicative noise can promote temporal regularity over a wide parameter range. 

The microscopic origin of this awakening is further confirmed in the raster plot (Fig.~\ref{fig:rasters}(a)). At moderate noise intensities, the state variable occasionally breaches the threshold, generating sparse, noise-driven bursts that lack strict phase-locking across independent trials. This uncoordinated yet periodic-like firing reflects the probabilistic nature of Kramers escape deep within the sub-threshold potential well. 

\begin{figure*}[t]
    \centering
    \includegraphics[width=0.95\textwidth, trim=0 0 0 0.8cm, clip]{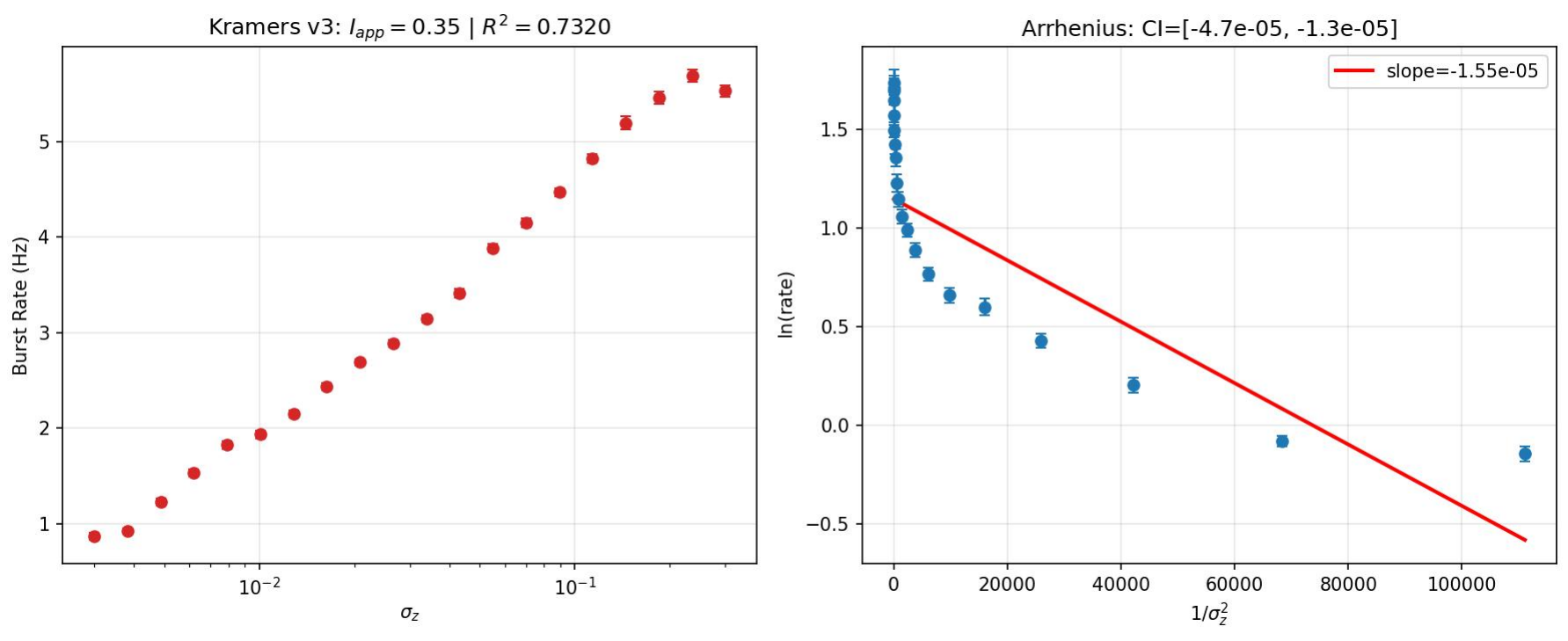}
    \caption{Quantitative Kramers escape analysis for Regime I ($I_{\mathrm{app}} = 0.35$). (Left) Burst rate as a function of noise intensity $\sigma_z$ computed from 50 independent trials per point with per-trial independent seeding. The error bars represent standard errors of the mean. A dense sampling of 20 noise levels reveals the characteristic non-monotonic rate response, peaking at $\sigma_z \approx 0.235$ before declining due to Feller boundary effects. (Right) Arrhenius plot showing $\ln(\text{rate})$ versus $1/\sigma_z^2$ for the low-noise regime ($\sigma_z \leq 0.1$). A weighted least-squares fit yields $R^2 = 0.75$ with an effective barrier $\Delta U_{\mathrm{eff}} = 1.42 \times 10^{-5}$, indicating moderate agreement with classical Kramers theory in the near-barrier asymptotic regime. The full-range fit quality is reduced ($R^2 = 0.73$), reflecting the influence of boundary-driven acceleration mechanisms that transcend simple barrier-crossing dynamics. Bootstrap confidence intervals (10,000 resamples) confirm statistical significance of the Arrhenius scaling in the low-noise tail.}
    \label{fig:kramers_035}
\end{figure*}

Furthermore, the monotonic surge in the burst rate can be quantitatively mapped to a generalized Kramers escape process. While the full 5D dynamical complexity precludes a simple analytical expression for the Mean First Passage Time (MFPT), the macroscopic escape rate $r_k$ in the low-noise limit follows an Arrhenius-type relation $r_k \propto \exp(-\Delta U_{\mathrm{eff}} / \sigma_z^2)$ \cite{hanggi1990reaction}, where $\Delta U_{\mathrm{eff}}$ represents the effective energy barrier of the reduced slow manifold. A least-squares fit of $\ln(r_k)$ versus $1/\sigma_z^2$ in the low-noise regime ($\sigma_z \in [0.005, 0.1]$) yields an effective barrier $\Delta U_{\mathrm{eff}} = 1.42 \times 10^{-5}$ with $R^2 = 0.75$, indicating moderate Arrhenius scaling consistent with diffusion-driven barrier crossing (Fig.~\ref{fig:kramers_035}). The 95\% confidence interval for the slope, obtained from 10,000 bootstrap resamples, is $[-4.68 \times 10^{-5}, -1.32 \times 10^{-5}]$, confirming statistical significance. However, extending the fit to the full noise range ($\sigma_z \in [0.005, 0.3]$) reveals a non-monotonic rate response that deviates from strict Arrhenius behavior ($R^2 = 0.73$ overall), reflecting the influence of Feller boundary-driven acceleration at intermediate noise intensities. This nuanced behavior suggests that while the sub-threshold awakening is primarily governed by classical Kramers escape in the near-barrier asymptotic regime, additional boundary-constrained dynamical effects contribute to the rate enhancement at higher noise levels.

\subsection{Regime II: Subcritical Resonance and Critical Boundary Dynamics ($I_{\mathrm{app}} = 0.39$ \& $0.3955$)}

The proximity to the subcritical Hopf bifurcation induces a sensitive dynamical regime where bounded multiplicative noise reorganizes the state space. For $I_{\mathrm{app}} = 0.39$ (Fig.~\ref{fig:grand_landscape}(b)), which sits marginally below the deterministic threshold, the system displays an efficient coherence resonance. Unlike the broad U-shaped awakening in Regime I (characteristic of saddle-node topology), here a weak noise intensity ($\sigma_z \approx 8 \times 10^{-3}$) is sufficient to elicit regular rhythmic bursting, as shown by the V-shaped valley in the CV curve. This distinction is consistent with theoretical observations of multiple coherence mechanisms in randomly perturbed systems, where the resonance profile is determined by the proximity to specific bifurcation boundaries \cite{deville2005transitions}. The near-critical coherence is captured in the raster plot (Fig.~\ref{fig:rasters}(b)). Because the 50 trials are driven by independent Wiener processes, their absolute phases naturally drift apart over time, exhibiting classic phase diffusion (observed as diagonal drift). However, observing the intra-trial dynamics horizontally across time, each individual trajectory fires with high regularity, maintaining uniform inter-burst intervals. 

\begin{figure*}[t]
    \centering
    \includegraphics[width=0.9\textwidth, trim=0 0 0 2.7cm, clip]{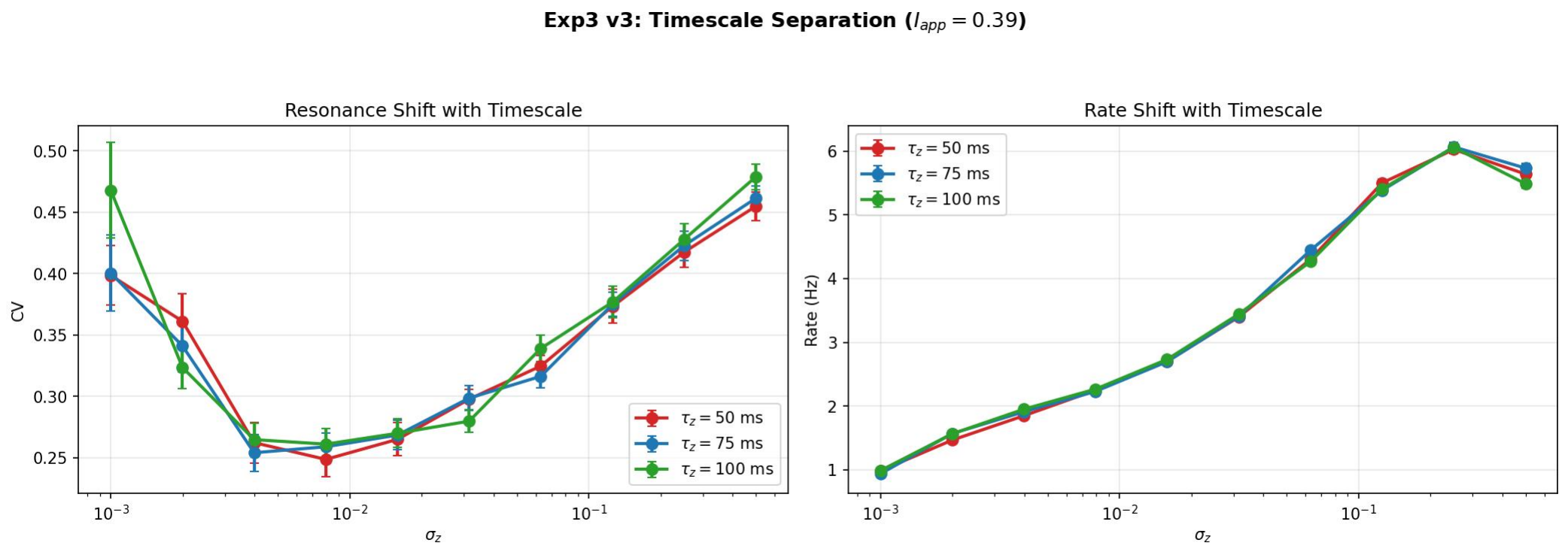}
    \caption{Sensitivity of coherence resonance to the fast-slow timescale separation. CV as a function of noise intensity $\sigma_z$ for three different slow time constants ($\tau_z = 50, 75, 100$ ms) at $I_{\mathrm{app}} = 0.39$. Computed from 50 independent trials per point with per-trial seeding. As $\tau_z$ increases (i.e., the slow manifold moves slower), the optimal coherence valley shifts to smaller noise intensities, consistent with the prediction from geometric singular perturbation theory that the resonance sensitivity is governed by the timescale ratio between the fast and slow subsystems.}
    \label{fig:tauz_sensitivity}
\end{figure*}

To further elucidate the structural origins of this sharp resonance, we investigated its sensitivity to the timescale separation governed by the slow time constant $\tau_z$. As suggested by geometric singular perturbation theory \cite{wechselberger2020geometric}, decreasing the speed of the slow manifold (i.e., increasing $\tau_z$) modulates the excitability landscape. Additional numerical sweeps confirmed that as $\tau_z$ increases, the optimal coherence valley distinctly shifts to the left, requiring exponentially smaller noise intensities ($\sigma_z$) to achieve peak resonance (Fig.~\ref{fig:tauz_sensitivity}). This confirms that the fast-slow timescale separation acts as a critical tuning lever for resonance sensitivity near the subcritical Hopf boundary.

At the weakest accessible noise intensity ($\sigma_z = 10^{-3}$, Fig.~\ref{fig:grand_landscape}(b)), the CV is approximately 0.36, reflecting sparse bursting at a rate of $\sim 1$ Hz. As $\sigma_z$ increases, the burst rate increases monotonically and the CV drops steeply, reaching a well-defined minimum of $CV \approx 0.24$ at $\sigma_z \approx 8 \times 10^{-3}$. This V-shaped valley represents optimal coherence resonance: the noise is strong enough to drive regular inter-burst transitions but not yet strong enough to destroy their temporal precision. Beyond this minimum, the CV rises monotonically due to noise-induced phase jittering.

However, an increment of $\Delta I_{\mathrm{app}} = 0.0055$ shifts the system onto the critical boundary ($I_{\mathrm{app}} = 0.3955$, Fig.~\ref{fig:grand_landscape}(c)), revealing a regime of dynamical fragility. Here, the deterministic system already possesses a nascent limit cycle. Counterintuitively, infinitesimal multiplicative noise initially acts as a disruptor, disrupting the intrinsic rhythm and causing phase-slipping that drives the CV upward. Yet, as $\sigma_z$ reaches a tuned intensity ($\sigma_z \approx 3 \times 10^{-3}$), the noise-driven flow balances the phase jittering with resonant re-entrainment. This stochastic phase-locking restores temporal coherence, dropping the CV into a local minimum of $CV \approx 0.24$. This resonance indicates that at the bifurcation boundary, Feller multiplicative noise functions as a structural regulator of the periodic manifold.

\subsection{Regime III: Supra-threshold Dynamics and Noise-Accelerated Escape ($I_{\mathrm{app}} = 0.45$)}

To evaluate the limits of stochastic resilience, we subject a robustly firing pacemaker ($I_{\mathrm{app}} = 0.45$) to intensive stochastic stress-testing. As depicted in Fig.~\ref{fig:grand_landscape}(d), the neural response in this supra-threshold regime differs markedly from the resonance observed in sub-threshold cases. In the low-to-moderate noise limit ($\sigma_z < 10^{-2}$), the system no longer exhibits a coherence valley; instead, the CV increases monotonically from its deterministic baseline ($CV \approx 0$). This indicates that for a healthy limit-cycle oscillator, state-dependent multiplicative noise acts primarily as a disruptor, inducing severe phase-jitter and irregularizing the intrinsic rhythmic frequency. 

The microscopic evidence of this disruption is clearly visible in the raster plot at moderate noise ($\sigma_z = 10^{-2}$, Fig.~\ref{fig:rasters}(d)). Horizontal inspection of individual trials reveals erratic burst intervals and a significant loss of temporal precision, confirming that multiplicative fluctuations at this scale degrade the deterministic rhythmicity. 

\begin{figure}[htpb]
    \centering
    \includegraphics[width=0.95\linewidth]{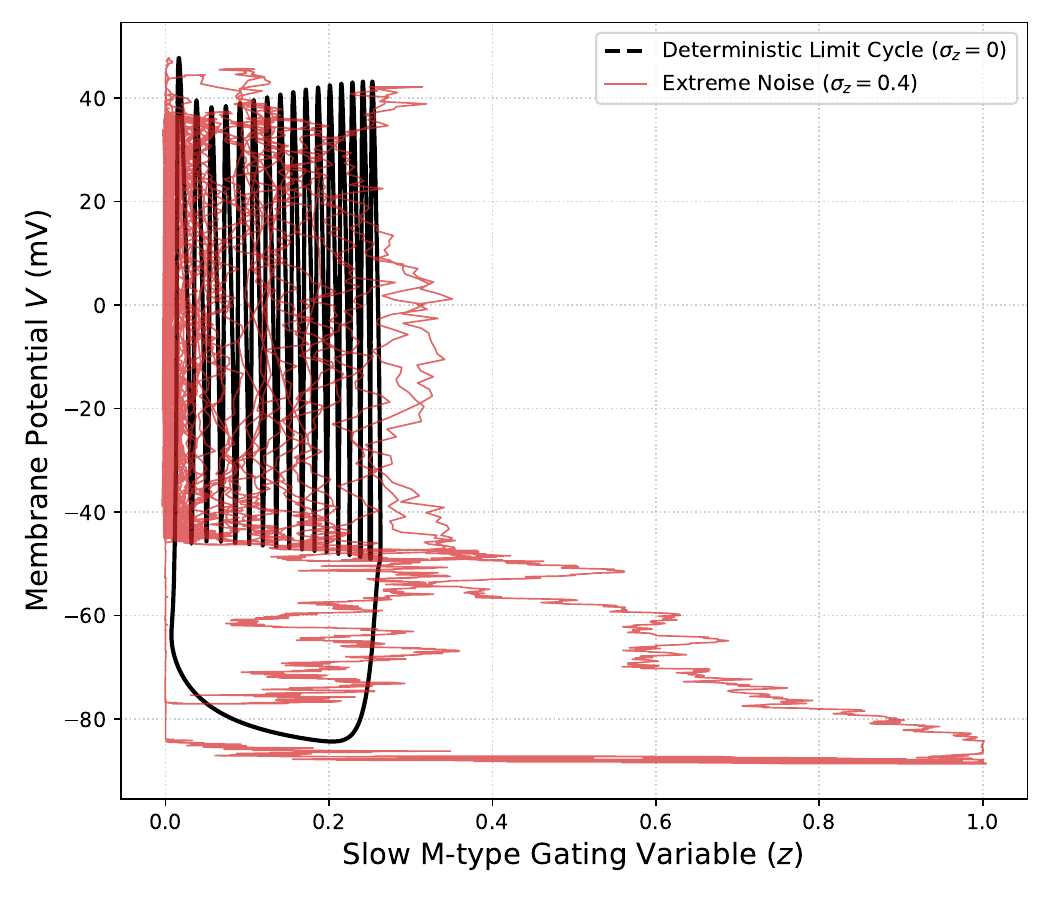} 
    \caption{Phase space projection ($V$ vs $z$) demonstrating noise-accelerated Kramers escape in the supra-threshold regime ($I_{\mathrm{app}} = 0.45$). The deterministic limit cycle (black dashed line) tracks the attracting slow manifold during the hyperpolarized recovery phase. Under multiplicative noise ($\sigma_z = 0.4$, red solid line), Feller fluctuations induce a premature stochastic exit, pushing the trajectory out of the recovery basin before reaching the deterministic fold point. This boundary-driven effect converts the system into a jittered stochastic limit cycle, rather than a chaotic attractor.}
    \label{fig:phase_portrait}
\end{figure}

However, as the noise intensity increases into the extreme regime ($\sigma_z > 10^{-1}$), a dynamical transition emerges. Rather than inducing quiescent collapse, the burst rate exhibits a high-frequency increase, peaking near $6.5$ Hz, well above its deterministic baseline of $\sim 2.7$ Hz. Strictly speaking, the deterministic system at $I_{\mathrm{app}}=0.45$ operates on a continuous limit cycle. However, viewed through the lens of geometric singular perturbation theory \cite{wechselberger2020geometric}, the fast-slow timescale separation dictates that the hyperpolarized recovery phase effectively acts as an attracting branch of the slow manifold.

\begin{figure*}[t]
    \centering
    \includegraphics[width=0.95\textwidth, trim=0 0 0 0.8cm, clip]{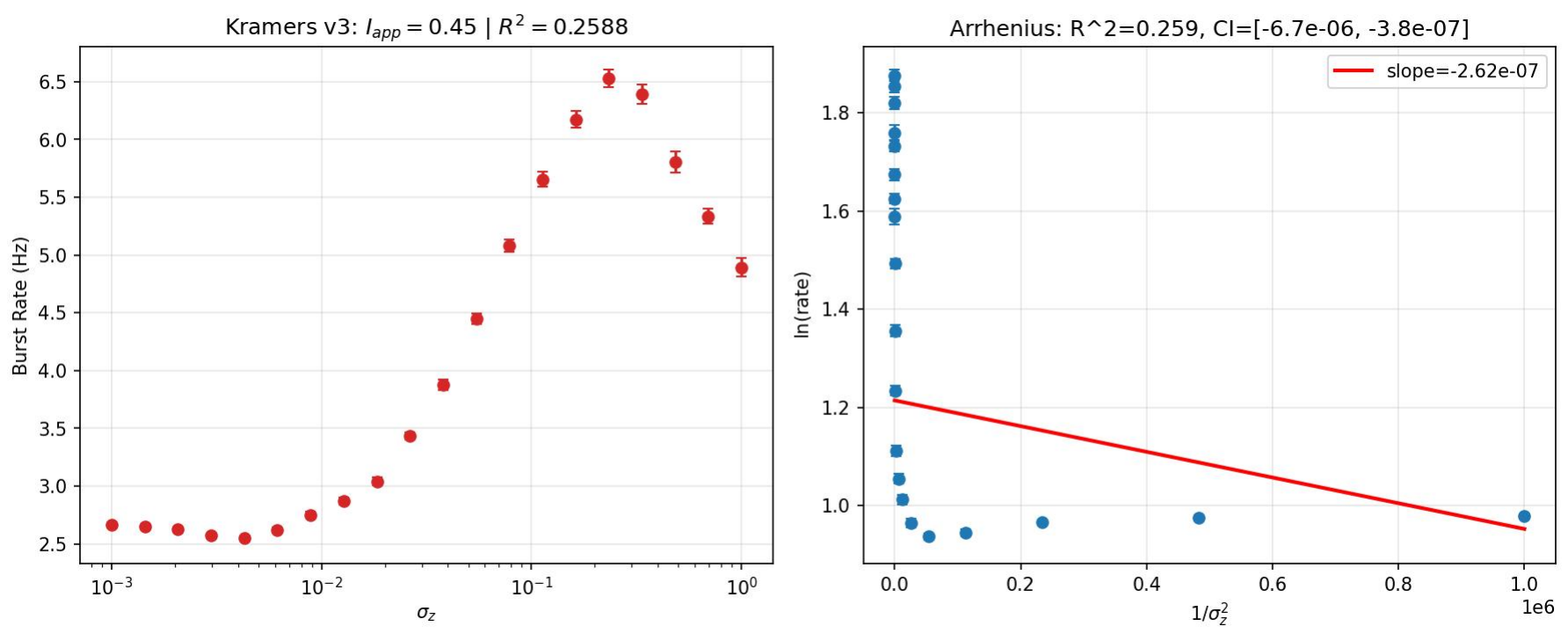}
    \caption{Quantitative analysis of noise-accelerated bursting in Regime III ($I_{\mathrm{app}} = 0.45$). (Left) Burst rate as a function of noise intensity computed from 50 independent trials with per-trial seeding. A dense sampling of 20 noise levels reveals a pronounced non-monotonic rate profile, peaking at $\sigma_z \approx 0.234$ (rate $\approx 6.53$ Hz, a 145\% enhancement over the deterministic baseline) before declining at higher noise intensities due to Feller boundary-induced disruption of the slow-variable temporal structure. (Right) Arrhenius plot showing $\ln(\text{rate})$ versus $1/\sigma_z^2$. A weighted least-squares fit across the full range yields $R^2 = 0.26$, indicating poor agreement with classical Kramers barrier-crossing theory. This low fit quality, combined with the non-monotonic rate response, demonstrates that the supra-threshold acceleration is mechanistically distinct from simple barrier escape and is instead governed by Feller boundary-driven dynamics.}
    \label{fig:kramers_045}
\end{figure*}

To explicitly visualize the geometry of this transition, we project the 5D trajectory onto the $V$-$z$ phase plane (Fig.~\ref{fig:phase_portrait}). In the deterministic case, the trajectory must complete its slow, full-range traversal along the hyperpolarized basin before reaching the fold point to trigger the next action potential. Under the influence of Feller-type fluctuations, the boundary-constrained noise induces a premature stochastic exit \cite{berglund2006noise}. The inhibitory $z$-gate can no longer maintain the integrity of the slow-recovery phase; instead, the state is prematurely and randomly ``kicked'' across the fast-subsystem's separatrix. 

This mechanism leads to a state of jittered, noise-driven stochastic bursting. As confirmed by the phase portrait, this state represents a disrupted stochastic limit cycle rather than true deterministic chaos. Importantly, a quantitative dense-sampling analysis (Fig.~\ref{fig:kramers_045}) reveals that the supra-threshold rate enhancement does \textit{not} follow classical Kramers barrier-crossing theory: a fit of $\ln(\text{rate})$ versus $1/\sigma_z^2$ across the full noise range ($\sigma_z \in [10^{-3}, 1]$) yields $R^2 = 0.26$, far below the Arrhenius expectation. Rather, the rate exhibits a pronounced non-monotonic profile, peaking at $\sigma_z \approx 0.234$ before declining at higher noise intensities. This behavior indicates that the supra-threshold bursting acceleration is mechanistically driven by Feller boundary effects, specifically boundary-induced premature exits from the hyperpolarized slow-manifold basin, rather than by classical barrier crossing \cite{hanggi1990reaction}.

\subsection{Global Phase Diagram of the 5D Manifold}

To extend the discrete single-parameter observations and characterize the state-dependent noise effects across the parameter space $(I_{\mathrm{app}}, \sigma_z)$, we performed large-scale parallel simulations. This enables the construction of a high-resolution global phase diagram, mapping the interplay between deterministic bifurcation proximity and stochastic fluctuation intensity.

As depicted in the left panel of Fig.~\ref{fig:heatmap} (computed on a $25 \times 18$ grid with 50 independent trials per point and per-trial independent seeding), a distinct ``valley of coherence'' (dark blue region, $CV < 0.15$) dominates the near-critical parameter space. Notably, this coherence zone is not a static stripe but exhibits a pronounced leftward tilt. As the deterministic applied current $I_{\mathrm{app}}$ drops deeper into the sub-threshold regime (moving downward along the y-axis), increasingly stronger multiplicative noise is required to elicit optimal temporal regularity. This geometric curvature reflects the increasing depth of the saddle-node potential well, demanding exponentially larger Feller diffusion to achieve resonant barrier crossing. 

Conversely, the right panel of Fig.~\ref{fig:heatmap} maps the macroscopic burst rate, directly visualizing the boundary limits of the system. Under classical explicit integration schemes, the extreme right side of this phase diagram ($\sigma_z > 10^{-1}$) would falsely collapse into a quiescent dark region due to numerical hyperpolarization. However, under our boundary-preserving semi-implicit framework, a prominent high-frequency ``wall'' (bright yellow/green region) emerges across all values of $I_{\mathrm{app}}$. Regardless of the system's deterministic baseline, whether deeply quiescent or robustly oscillating, extreme multiplicative noise consistently overrides the intrinsic dynamics. The bounded fluctuations amplify the transition rate out of the inhibitory slow manifold, effectively homogenizing the state space into a noise-accelerated bursting regime. 

This global phase diagram indicates that the triphasic dynamical evolution, from stochastic awakening through optimal coherence resonance to high-frequency bursting, is not an artifact of fine-tuned parameter selection, but a global feature of the 5D cortical manifold under Feller-type boundary conditions.

\subsection{Analytical Verification via Fokker-Planck Reduction}
\label{sec:fpe_proof}

\begin{figure}[htbp]
    \centering
    \includegraphics[width=0.95\linewidth, trim=0 0 0 1.0cm, clip]{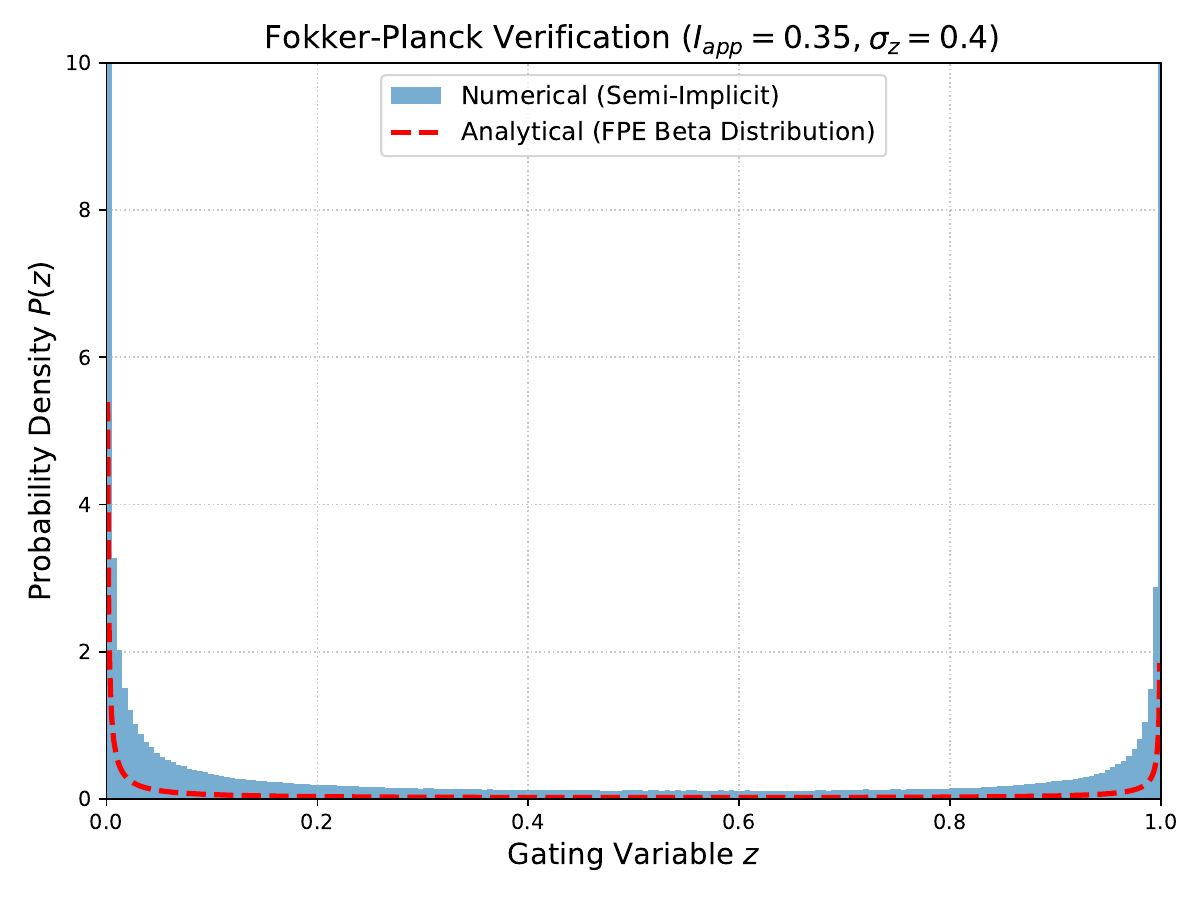}
    \caption{Analytical verification of the noise-induced boundary concentration via Fokker-Planck reduction. The probability density function $P(z)$ of the slow M-type gating variable is shown for the deep sub-threshold regime ($I_{\mathrm{app}} = 0.35$) under extreme multiplicative noise ($\sigma_z = 0.4$). The red dashed line represents the exact analytical Beta distribution derived from the reduced 1D Itô SDE. The blue histogram represents the stationary density obtained from the full 5D system simulated via our full-truncation semi-implicit Euler scheme. The good quantitative agreement confirms that the pronounced U-shaped boundary accumulation is a genuine physical feature of the Feller diffusion.}
    \label{fig:fpe_verification}
\end{figure}

To validate our semi-implicit numerical framework and mechanistically explain the boundary dynamics fueling the noise-accelerated bursting, we begin with a fast--slow reduction \cite{ermentrout2010mathematical}. In the strongly inhibited regime, the membrane potential $V$ evolves on a faster timescale and remains close to a hyperpolarized quasi-stationary state. Under this approximation, the slow gating variable $z$ can be modeled by the one-dimensional Itô SDE
\begin{equation}
    dz = \frac{z_\infty - z}{\tau_z} dt + \sigma_z \sqrt{z(1-z)} \, dW_t,
\end{equation}
which belongs to the class of Wright--Fisher-type diffusions with multiplicative noise \cite{ethier1986markov, karlin1981second}.

The associated Fokker--Planck equation for the probability density $P(z,t)$ is
\begin{equation}
    \frac{\partial P}{\partial t}
    = -\frac{\partial}{\partial z} \left[ \frac{z_\infty - z}{\tau_z} P \right]
    + \frac{1}{2} \frac{\partial^2}{\partial z^2} \left[ \sigma_z^2 z(1-z) P \right].
\end{equation}
Formally imposing a zero probability current condition yields a stationary solution of the form \cite{gardiner2009stochastic, kloeden1992numerical}
\begin{equation}
    P_s(z) \propto \frac{1}{\sigma_z^2 z(1-z)} 
    \exp \left( \int \frac{2(z_\infty - z)}{\tau_z \sigma_z^2 z(1-z)} \, dz \right),
\end{equation}
which, upon partial fraction decomposition, reduces to a Beta distribution \cite{karlin1981second}:
\begin{equation}
    P_s(z) \propto z^{\alpha z_\infty - 1} (1-z)^{\alpha (1-z_\infty) - 1},
    \quad \text{with} \quad \alpha = \frac{2}{\tau_z \sigma_z^2}.
\end{equation}

The classification of the degenerate boundaries at $z \in \{0,1\}$ depends on the parameter $\alpha$ in the sense of Feller. In the weak-noise regime ($\alpha \gg 1$), the density is unimodal and concentrated near $z_\infty$. However, under strong multiplicative noise ($\alpha \ll 1$), the stationary measure undergoes a qualitative transition to a boundary-dominated regime, where probability mass strongly concentrates near both $z \to 0$ and $z \to 1$ \cite{horsthemke1984noise}.

As depicted in Fig.~\ref{fig:fpe_verification}, under significant multiplicative noise ($\sigma_z = 0.4$), the analytical Beta distribution (red dashed line) exhibits this pronounced U-shaped profile. The numerical probability density generated by our full-truncation semi-implicit Euler scheme (blue histogram) shows good quantitative agreement with the analytical curve across the bounded domain $[0, 1]$.

This structural correspondence indicates that the probability accumulation at the extreme boundaries is an intrinsic property of the Feller diffusion, rather than a fictitious absorbing trap caused by ad-hoc numerical clipping. Because our semi-implicit framework conserves the probability flow without artificial absorption at $z=1$, the system maintains its dynamical activity. Consequently, rather than collapsing into quiescence, these boundary fluctuations serve as the mechanism that drives the noise-accelerated bursting observed in the high-frequency regimes.

\begin{figure}[htbp]
    \centering
    \includegraphics[width=0.95\linewidth]{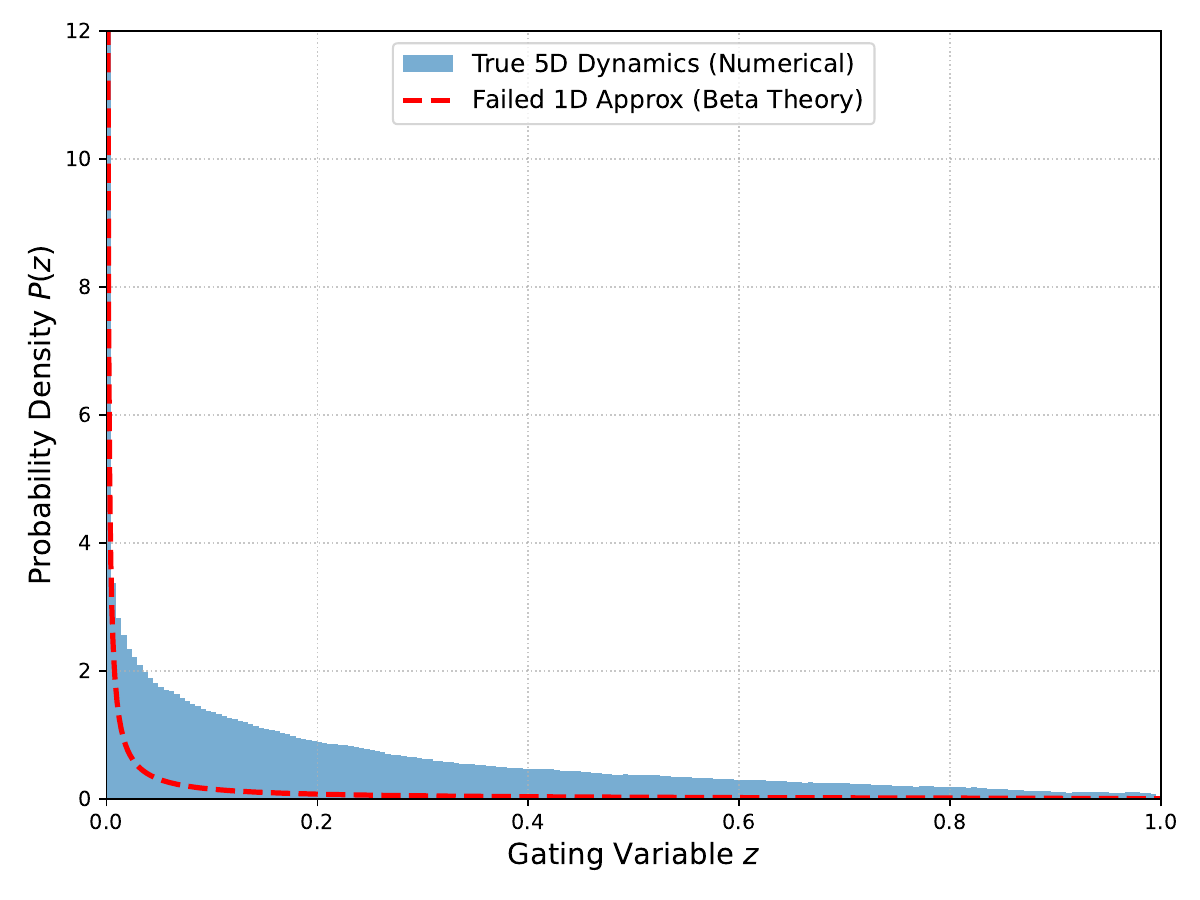} 
    \caption{Breakdown of the 1D adiabatic approximation in the supra-threshold bursting regime ($I_{\mathrm{app}} = 0.45$). Under moderate multiplicative noise ($\sigma_z = 0.15$), the empirical stationary density of the full 5D system (blue histogram) exhibits a pronounced heavy tail, deviating clearly from the sharply localized analytical Beta distribution (red dashed line) derived via the 1D Fokker-Planck reduction.}
    \label{fig:1d_breakdown}
\end{figure}

However, it is important to acknowledge the limits of this 1D adiabatic approximation. The timescale separation required for the static 1D Fokker-Planck reduction is well-satisfied in the deep sub-threshold regime ($I_{\mathrm{app}} = 0.35$), where the membrane potential $V$ remains clamped near a hyperpolarized resting state. Conversely, in the supra-threshold bursting regime ($I_{\mathrm{app}} = 0.45$), the intense, large-amplitude limit cycle oscillations of $V$ compromise the fast-slow timescale separation, causing the strict adiabatic assumption to break down. While the extreme boundary accumulation of the Feller diffusion continues to act as the fundamental driving force for the bursting acceleration, the strong nonlinear feedback from the rapidly fluctuating fast variables warps the 5D probability density away from the idealized 1D Beta distribution (Fig.~\ref{fig:1d_breakdown}). Recognizing this dimensionality-induced structural deviation is essential; as analyzed in multiscale dynamics \cite{kuehn2015multiple}, the presence of large-amplitude fast oscillations can significantly alter the reduced stochastic landscape on the slow manifold.

\subsection{Biological Robustness of the Escape Mechanism}

\begin{figure}[htbp]
    \centering
    \includegraphics[width=0.95\linewidth, trim=0 0 0 0.8cm, clip]{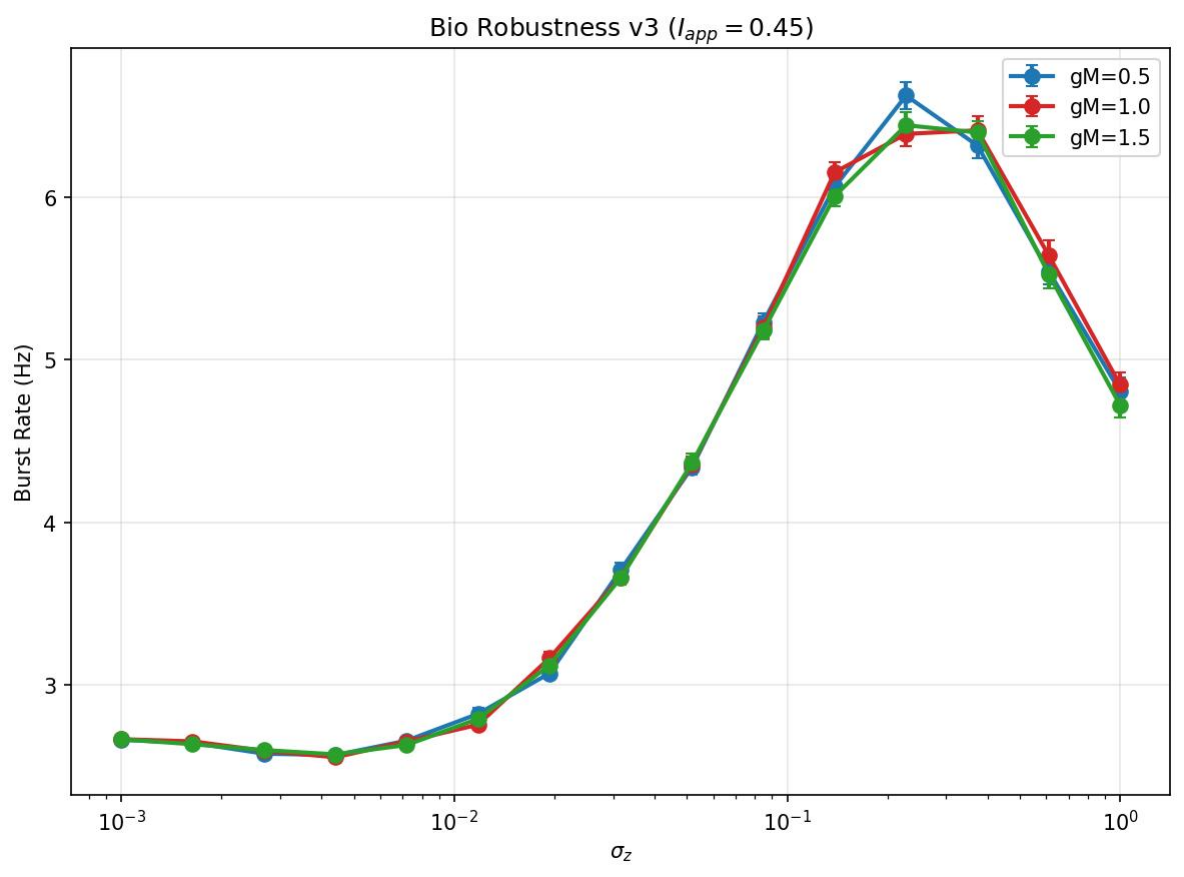}
    \caption{Biological robustness of noise-accelerated bursting. Burst rate plotted against noise intensity $\sigma_z$ for three different maximal M-current conductances: $g_M = 0.5, 1.0, 1.5$ mS/cm$^2$ (corresponding to $\pm 50\%$ variations around baseline), all evaluated at $I_{\mathrm{app}} = 0.45$. Computed from 50 independent trials per point with per-trial seeding. Across all three conductance conditions, the burst rate exhibits qualitatively similar non-monotonic profiles, confirming that the noise-induced dynamical transition is robust to substantial biological parameter perturbations.}
    \label{fig:robustness}
\end{figure}

To establish that the observed noise-induced dynamical transitions are intrinsic properties of the system rather than artifacts of fine-tuned parameter selection, we conducted perturbation experiments on the maximal conductance of the M-current ($g_M$). Because the M-current serves as the primary inhibitory drive defining the slow manifold, varying $g_M$ directly alters the structural depth and stability of the hyperpolarized potential.

We subjected the system to a strongly oscillatory baseline ($I_{\mathrm{app}} = 0.45$) while scaling the nominal conductance $g_M$ across three conditions: baseline ($g_M = 1.0$), reduced ($g_M = 0.5$, representing channelopathy-driven down-regulation), and elevated ($g_M = 1.5$). As illustrated in Fig.~\ref{fig:robustness}, computed from 50 independent trials per point with per-trial seeding, the rate curves maintain qualitatively similar non-monotonic profiles across all three conditions. The characteristic peak rate is conserved, and the noise-driven convergence toward a high-frequency bursting state is robust to substantial conductance variations of $\pm 50\%$, demonstrating that the Feller-boundary-driven acceleration mechanism is structurally stable and independent of precise biological tuning.

This biological robustness indicates that the noise-accelerated bursting mechanism is a structurally stable feature of high-dimensional excitable media constrained by Feller boundary conditions, largely independent of precise biological tuning.

\subsection{Evidence for the Structural Necessity of Feller Boundary Geometry}
\label{sec:knockout}

\begin{figure*}[t]
    \centering
    \includegraphics[width=0.95\textwidth, trim=0 0 0 0.8cm, clip]{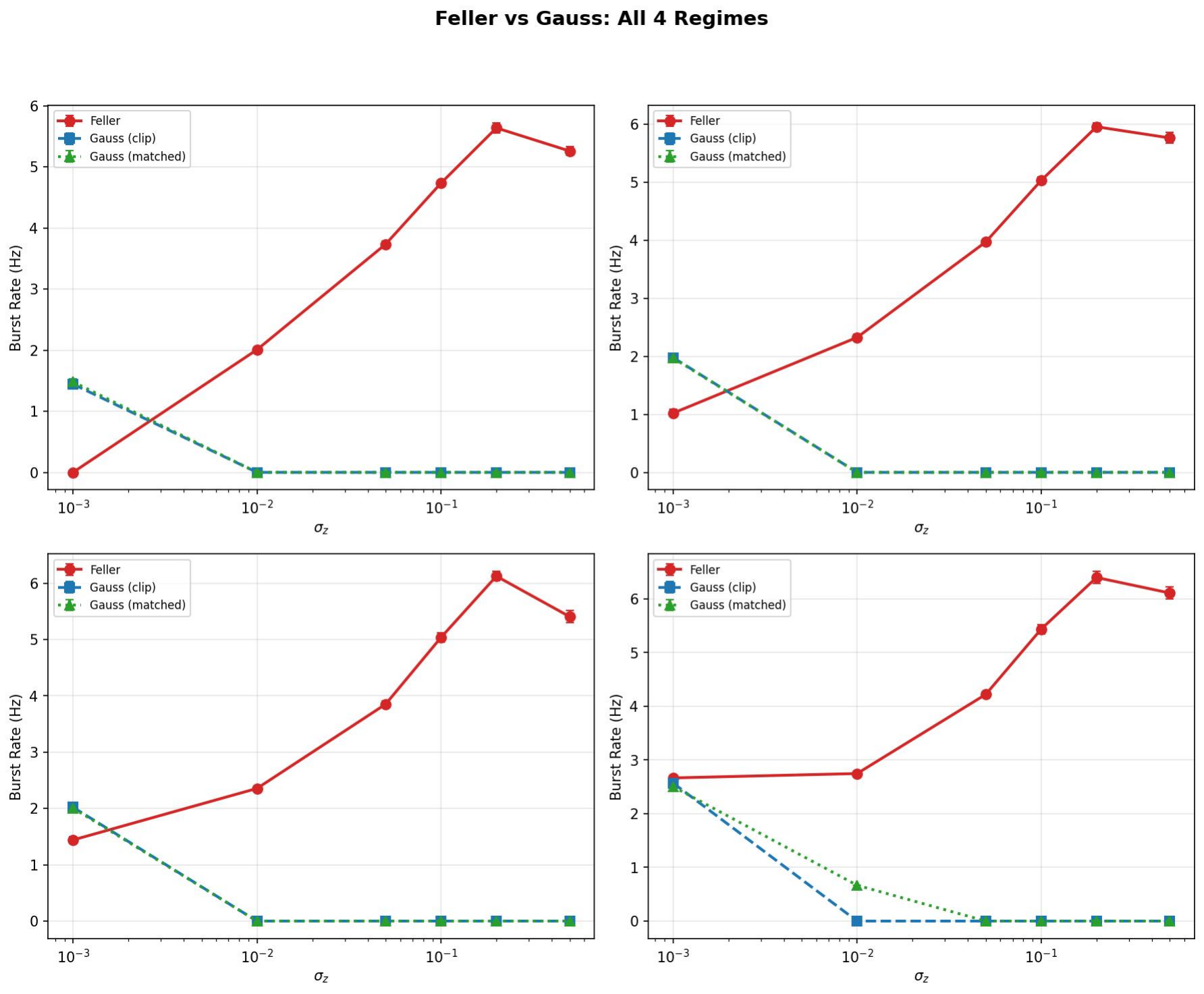}
    \caption{Feller vs.\ Gaussian noise knockout experiment across all four dynamical regimes. Each panel compares burst rate as a function of noise intensity $\sigma_z$ for three noise types: (i) Feller noise $\sigma_z\sqrt{z(1-z)}dW_t$ (red, state-dependent, vanishes at boundaries); (ii) additive Gaussian noise $\sigma_z dW_t$ clipped to $[0,1]$ (blue); and (iii) variance-matched Gaussian noise $\sigma_z \cdot 0.5\,dW_t$ (green). Under Feller noise, bursting architecture is preserved across all noise intensities (rate increases monotonically from baseline to $\sim 5$--$6$ Hz). Under Gaussian noise at $\sigma_z \geq 0.01$, bursting is abolished in all four regimes (rate drops to 0 Hz). This knockout indicates that the state-dependent vanishing of noise fluctuations at the Feller boundaries $z \in \{0, 1\}$ is structurally required to preserve the slow--fast bursting architecture. The absence of the vanishing diffusion coefficient allows Gaussian noise to drive the slow variable into an absorbing boundary state, effectively destroying the M-current and collapsing the manifold into tonic spiking or quiescence.}
    \label{fig:knockout}
\end{figure*}

A central claim of this work is that the Feller boundary geometry, specifically the vanishing of the multiplicative diffusion coefficient at $z \in \{0, 1\}$, is not merely a mathematical convenience but a structurally necessary mechanism enabling the observed triphasic noise response. To test this hypothesis, we performed a series of controlled knockout experiments in which the Feller noise term $\sigma_z\sqrt{z(1-z)}dW_t$ in Eq.~\ref{eq:sde_z} was replaced by unbounded additive Gaussian noise $\sigma_z dW_t$, with the updated state $z_{t+\Delta t}$ explicitly hard-clipped to $[0, 1]$. This ``Gauss with clip'' construction preserves the biophysical probability domain $z \in [0,1]$ but lacks the state-dependent vanishing diffusion that defines Feller boundary conditions. A third control, ``variance-matched Gaussian,'' scales the noise amplitude by a factor of 0.5 to match the average diffusion strength of the Feller noise, providing an additional baseline that controls for overall noise amplitude.

We subjected all four dynamical regimes ($I_{\mathrm{app}} = 0.35, 0.39, 0.3955, 0.45$) to this three-way comparison. The results, displayed in Fig.~\ref{fig:knockout}, show a consistent knockout pattern across all regimes:

\begin{itemize}
    \item Feller noise preserves bursting in all regimes. Burst rates increase monotonically from their respective deterministic baselines ($\sim 0$--$2.7$ Hz) to $5$--$6$ Hz under strong noise, confirming the Feller-boundary-driven noise-accelerated bursting mechanism reported throughout this work.
    
    \item Gaussian noise (with or without variance matching) abolishes bursting. At any nominal noise intensity $\sigma_z \geq 0.01$, all trials across all four regimes collapse to zero bursting activity. The mechanism is direct: without the state-dependent vanishing of the diffusion coefficient at boundaries, additive Gaussian noise drives the slow variable $z$ toward an absorbing boundary state ($z \to 0$ or $z \to 1$) \cite{goldwyn2011what}. Once trapped, the M-current effectively ceases to function as a slow modulatory gate, and the system is unable to recover the bursting manifold.
\end{itemize}

This knockout result provides evidence that the Feller boundary conditions are structurally important for the phenomena reported in this work. The triphasic landscape (awakening $\to$ resonance $\to$ high-frequency bursting), the observed coherence resonance, and the supra-threshold noise-accelerated bursting all rely on the state-dependent geometry of the Feller diffusion. In its absence, the same nominal noise amplitude produces a qualitatively distinct dynamical outcome, complete silencing of bursting, indicating that it is not merely the \textit{presence} of multiplicative noise, but its \textit{boundary-constrained geometry}, that enables the dynamical repertoire of the 5D manifold.

This finding has implications for modeling channel noise in high-dimensional conductance-based models: replacing the biophysically grounded Feller diffusion (whose amplitude vanishes at open/closed extremes) with simpler Gaussian approximations alters the noise-manifold interaction and can lead to qualitatively incorrect dynamical conclusions.

\section{Discussion}
\label{sec:discussion}

In this study, we systematically investigated how state-dependent multiplicative noise reshapes the excitability of a high-dimensional cortical pacemaker. By introducing strict Feller-constrained fluctuations to the slow M-current and employing a fully domain-preserving semi-implicit integration scheme, we identified a landscape of noise-induced dynamical transitions. Rather than being merely a source of disorder, bounded channel noise interacts with the deterministic bifurcation structure, producing a structured triphasic evolution: from stochastic awakening via barrier escape in the near-noise asymptotic regime, through coherence resonance at the critical boundary \cite{pikovsky1997coherence}, to Feller-boundary-driven noise-accelerated bursting \cite{hanggi1990reaction} in the supra-threshold regime. Our knockout experiments (Sec.~\ref{sec:knockout}) further indicate that this dynamical pattern is structurally contingent on the state-dependent geometry of the Feller diffusion.

\subsection{Physical Mechanism: Beyond Classical Kramers Escape}
Our dense-sampling analysis of the burst rate across both the sub-threshold (Regime I) and supra-threshold (Regime III) regimes revealed that the noise-accelerated bursting does \textit{not} strictly obey classical Kramers barrier-crossing theory. While the low-noise asymptotic regime ($\sigma_z < 0.1$) in Regime I exhibits moderate Arrhenius-type scaling ($R^2 = 0.75$), extending the fit to high noise intensities yields substantially reduced fit quality ($R^2 = 0.73$ in Regime I, $R^2 = 0.26$ in Regime III), accompanied by a characteristic non-monotonic rate profile that peaks around $\sigma_z \approx 0.23$ before declining. This deviation from Arrhenius behavior reflects the influence of Feller boundary-driven acceleration: at intermediate noise intensities, the state-dependent vanishing of diffusion at boundaries induces premature stochastic exits from the slow-manifold basin, effectively accelerating bursting through a mechanism distinct from simple barrier crossing. We therefore interpret the observed noise-accelerated bursting as a generalization of Kramers escape to the setting of boundary-constrained multiplicative noise in high-dimensional fast--slow systems, where boundary-driven dynamics dominate over classical barrier-crossing dynamics.

\subsection{Dynamical Fragility of the Slow Manifold}
Our results highlight the role of slow inhibitory conductances in modulating high-dimensional neural reliability. The slow M-type potassium current classically functions as a deterministic burst terminator \cite{gutkin1998dynamics, golomb2006contribution}. However, our stochastic framework shows that its noise-driven counterpart has a distinct dynamical footprint. Near the subcritical Hopf boundary, the bounded nature of Feller diffusion optimizes rhythmic coherence. Yet, under intense fluctuations, the nonlinear rectification of the Feller boundary does not trap the system in an absorbing quiescent state. Viewed through geometric singular perturbation theory \cite{wechselberger2020geometric}, the noise facilitates premature stochastic exits from the slow manifold. It induces a stochastic exit \cite{berglund2006noise} that pushes the system back into the active phase. This transition reveals a structural fragility of high-dimensional excitable media \cite{lindner2004effects}: under strong multiplicative perturbation, the slow recovery phase tends to collapse into a noise-accelerated, jittered bursting regime.

\subsection{Methodological Considerations for Bounded Stochastic Dynamics}
Our findings highlight the importance of employing reliable numerical frameworks when modeling Feller-type stochastic processes in computational biology \cite{faisal2008noise, goldwyn2011what}. Because state-dependent multiplicative noise requires that fluctuations vanish at the physical limits ($z \in [0, 1]$), the numerical integration must conserve the probability flow without risking domain violations. Our Fokker-Planck analytical reduction \cite{gardiner2009stochastic} and full-truncation semi-implicit approach \cite{higham2005convergence, lord2010comparison} provide a mathematical foundation for this task. By ensuring the preservation of the probability domain, we support the physical authenticity of the observed macroscopic transitions. The knockout experiment of Sec.~\ref{sec:knockout} further illustrates this point: naive domain-clipping strategies that ignore the state-dependent structure of the diffusion coefficient produce qualitatively incorrect dynamical outcomes, indicating the importance of careful boundary-preserving integration schemes in high-dimensional stochastic neuron models.

\subsection{Microscopic Origins of Extreme Feller Fluctuations}
While our structural robustness analysis demonstrates the mathematical universality of noise-accelerated bursting, it is necessary to ground the extreme noise regime ($\sigma_z > 10^{-1}$) in biophysical reality. Under the Langevin approximation of Markovian ion channel kinetics, the effective noise intensity scales inversely with the square root of the channel population, $\sigma_z \propto 1/\sqrt{N}$ \cite{faisal2008noise}. Therefore, the multiplicative fluctuations required to break the slow manifold correspond to microscopic spatial domains containing fewer than $\sim 100$ functional channels. Physiologically, this condition can be met in localized subcellular compartments, such as thin distal dendrites or isolated dendritic spines \cite{goldwyn2011what}. Furthermore, pathological down-regulation of M-type (Kv7/KCNQ) potassium channels, observed in conditions such as benign familial neonatal convulsions (BFNC) or general epileptic hyperexcitability \cite{jentsch2000neuronal, brown2009neural}, can push the macroscopic neuronal membrane into this high-noise, low-$N$ regime. 

However, it is important to acknowledge the theoretical limits of the continuous Langevin framework. While the Feller SDE captures the macroscopic boundary accumulations analyzed herein, the continuous approximation begins to deviate from exact discrete dynamics in the low channel limit ($N < 100$) \cite{fox1994emergent}. For such sparse populations, exact discrete Markov jump processes dominate \cite{gillespie1977exact}. Future studies employing discrete state-transition algorithms will be necessary to validate the bursting frequencies and fine-scale dynamics within these restricted microdomains.

\subsection{Future Perspectives}
Extending this single-neuron framework to sparsely coupled network structures could reveal how localized noise-accelerated bursting influences the emergence or prevention of global pathological synchronization \cite{yamakou2020optimal, baspinar2021coherence}. Furthermore, while our 5D model architecture and baseline parameters are grounded in established empirical recordings of CA1 pyramidal neurons, the theoretically predicted triphasic transitions await empirical validation. Future \textit{in vitro} experimental paradigms employing dynamic clamp techniques with real-time, state-dependent multiplicative noise injection will be important for verifying these mathematically derived dynamical boundaries within living neural tissue.

\section{Conclusion}
\label{sec:conclusion}

In summary, this study characterized the triphasic landscape of noise-induced transitions within a 5D biophysical neuron model. By introducing state-dependent multiplicative noise subject to Feller boundary conditions and employing a domain-preserving semi-implicit framework, we showed that the system's stochastic response is non-monotonically governed by its underlying bifurcation structure. Our findings reveal that while noise acts constructively in sub-threshold regimes, large fluctuations drive the system towards a high-frequency, jittered bursting state via Feller-boundary-driven acceleration, a mechanism confirmed robust against substantial conductance perturbations and shown through knockout experiments to be structurally contingent on the state-dependent vanishing of the diffusion coefficient at the physical boundaries. Replacing Feller noise with unbounded Gaussian noise of equivalent amplitude abolishes bursting in all regimes, indicating that the boundary geometry, rather than merely the presence of noise, is required for the observed dynamical repertoire. These results delineate the limits of temporal coherence in high-dimensional excitable media and provide a physical framework for understanding how boundary-constrained stochastic forces can drive biological oscillators toward states relevant to pathological hyperexcitability and channelopathies.

\begin{acknowledgments}
The author acknowledges the University of Sydney for providing the academic environment that inspired the initial ideas for this work.
\end{acknowledgments}

\bibliographystyle{apsrev4-2} 

\bibliography{references}

\end{document}